\documentclass[]{aastex631}

\submitjournal{ApJ}

\usepackage{graphicx}	 
\usepackage{amsmath}	 
\usepackage{amssymb}	 
\usepackage{mathtools}   
\usepackage{CJK}         
\usepackage{xcolor}      
\usepackage{hyperref}
\hypersetup{
    colorlinks=true,
    linkcolor=blue,
    filecolor=magenta,      
    urlcolor=blue,
    }

\begin{document}
\begin{CJK*}{UTF8}{gbsn}

\title[MHD Simulations of Disk Eccentricity Growth]{Simulations of Eccentricity Growth in Compact Binary Accretion Disks with MHD Turbulence}

\author{Morgan Ohana}
\affiliation{Department of Physics, University of California, Santa Barbara, CA 93106, USA}

\author[0000-0002-2624-3399]{Yan-Fei Jiang (姜燕飞)}
\affiliation{Center for Computational Astrophysics, Flatiron Institute, New York, NY 10010, USA}

\author[0000-0002-8082-4573]{Omer Blaes}
\affiliation{Department of Physics, University of California, Santa Barbara, CA 93106, USA}

\author{Bryance Oyang}
\affiliation{Department of Physics, University of California, Santa Barbara, CA 93106, USA}



\begin{abstract}

We present the results of four magnetohydrodynamic simulations and one alpha-disk simulation of accretion disks in a compact binary system, neglecting vertical stratification and assuming a locally isothermal equation of state. We demonstrate that in the presence of net vertical field, disks that extend out to the 3:1 mean motion resonance grow eccentricity in full MHD in much the same way as in hydrodynamical disks.  
Hence turbulence due to the magnetorotational instability (MRI) does not impede the tidally-driven growth of eccentricity in any meaningful way.  However, we find two important differences with alpha-disk theory.  First, in MHD, eccentricity builds up in the inner disk with a series of episodes of radial disk breaking into two misaligned eccentric disks, separated by a region of circular orbits.  Standing eccentric waves are often present in the inner eccentric disk.   Second, the successful spreading of an accretion disk with MRI turbulence out to the resonant radius is nontrivial, and much harder than spreading an alpha-disk.  This is due to the tendency to develop over-dense rings in which tidal torques overwhelm MRI transport and truncate the disk too early. We believe that the inability to spread the disk sufficiently was the reason why our previous attempt to excite eccentricity via the 3:1 mean motion resonance with MHD failed. 
Exactly how MHD disks successfully spread outward in compact binary systems is an important problem that has not yet been understood.


\end{abstract}

\keywords{Stellar accretion disks(1579) --- Magnetohydrodynamics(1964) --- Cataclysmic variable stars(203)}




\section{Introduction}
\label{sec:introduction}

The luminosity observed from accretion disks formed by mass transfer in binary systems commonly exhibits distinct classes of variability on a range of time scales, from stochastic variability to periodic and quasiperiodic behavior to transient outbursts.  Superhumps are a particular type of periodic behavior that occur in sufficiently compact, circular binaries.  First observed in SU~UMa-type dwarf novae in outburst \citep{vogt74,war75}, they also exist in some persistent accreting white dwarf systems \citep{pat93} as well as some X-ray binaries \citep{odo96,zur02}.  Superhumps generally occur at periods slightly longer (positive superhumps) or slightly shorter (negative superhumps) than the orbital period of the binary, and are believed to be associated with prograde apsidal precession of an eccentric disk in the former case, or retrograde nodal precession of an inclined disk in the latter case.  We wish to focus on positive superhumps in this paper. 

The origin of disk eccentricity in those systems that exhibit positive superhumps appears to be inextricably tied to a mean motion resonance within the accretion disk, where the material orbits the accretor star three times every binary orbital period.  A theoretical mechanism for how this resonance drives eccentricity growth throughout the disk was elaborated by \citet{lubow_theory}.  A perturbative eccentricity in the disk interacts nonlinearly with the nonaxisymmetric tidal field of the companion star (particularly the $m=3$ azimuthal Fourier component) to launch a wave in the disk from the 3:1 mean motion resonance.  This wave, in turn, interacts nonlinearly with the tidal field to produce a positive feedback on the eccentricity.

Numerical simulations of the tidally-driven growth of eccentricity in accretion disks in compact binaries were first successfully performed using viscous smooth particle hydrodynamics \citep{whi88,whi91,whi94,kun97,mur98}.  Eccentricity growth in grid-based viscous hydrodynamic simulations proved to be more difficult to achieve, but was first robustly observed by \citet{kle08}, who found consistency with the mechanism proposed by \citet{lubow_theory}.  However, the viscous angular momentum transport used in all these simulations is necessarily artificial, as it really represents an effective transport due to turbulence that is thought to be present in these flows.  How such turbulence interacts with the waves thought to be responsible for the eccentricity growth, and indeed the eccentricity itself, are questions that have only recently received attention.

Given that the systems that exhibit superhumps should be in a regime where the accretion disks are highly ionized and electrically conducting, a robust mechanism for driving turbulence is the magnetorotational instability (MRI; \citealt{bal98}).  Several groups have now conducted MHD simulations of eccentric disks with MRI turbulence.  \citet{dew20} conducted global, vertically unstratified simulations of disks in a pseudo-Newtonian potential with a forced eccentricity applied at the outer radial boundary.  They found that sufficiently strong eccentricity forcing excited trapped inertial waves in this potential, overcoming the damping of such waves by MRI turbulence.  Indeed, they even found that eccentricity forcing could reduce the Maxwell stress associated with MRI turbulence.  \citet{lyn23} examined the effects of vertical magnetic fields on eccentric modes in 2D simulations, though without the development of the MRI.  Following a linear analysis of the MRI in eccentric disks \citep{chan18}, \citet{chan22} conducted global general relativistic MHD simulations of disks with eccentricity imposed in the initial condition.  These simulations again neglected vertical stratification, and used a nearly-Newtonian metric with a potential designed to have no apsidal precession of eccentric orbits, replicating the closed-orbit nature of bound particle trajectories  that occur for a spherical mass in Newtonian mechanics.  They found that the plasma beta and effective alpha parameter \citep{sha73} due to Maxwell stresses saturated at comparable levels to that seen in circular disk simulations, with the same dependence on magnetic topology.  But they also found that alpha changes sign with azimuth, and that MRI turbulence is more effective at outward transport of angular momentum than energy for eccentric disks.  This resulted in the growth of low angular momentum eccentric voids in the inner disk where periapsis coincides with their inner boundary.  \citet{chan23} extended this work to include vertical stratification.  The resulting azimuthally-dependent vertical compression near periapsis enhanced the rate of magnetic dissipation in their simulations.  Vertical stratification also enhanced the rate at which MRI turbulence alters the radial eccentricity profile.  

The eccentricity observed in all these MHD simulations arose from direct forcing or from imposing it in the initial conditions.  In \citet{Oyang2021}, we investigated the growth of eccentricity in a binary system in the presence of MRI turbulence.  We conducted vertically stratified MHD simulations in the binary potential, with a magnetic field fed in by the accretion stream that had no net magnetic flux, leading to an effective $\alpha$ parameter of 0.01. The disk failed to spread outward sufficiently against the effects of tidal truncation to produce significant surface density at the 3:1 resonance.  This then stymied the Lubow mechanism, resulting in negligible disk eccentricity.
Stronger MRI turbulent stresses therefore appeared to be necessary to spread the disk to the resonance radius and grow eccentricity.

In this paper we follow the work of \citet{Oyang2021} by conducting MHD simulations in a binary potential with net poloidal flux in order to strengthen the MRI stresses.  In contrast to \citet{Oyang2021}, we neglect vertical stratification here in order to avoid complications associated with MHD outflows that are characteristic of vertically stratified simulations with net poloidal flux (e.g. \citealt{suz09,fro13}).  We find that disks that are initialized or form within the resonance radius can face challenges in spreading outward to that radius, regardless of magnetic field topology.
This is because surface density can build up if the outward spreading stalls, and tidal torques (which are proportional to density) then become more effective at extracting angular momentum than MRI Maxwell stresses can resupply it, driving the disk to truncate.  On the other hand, disks that are initialized beyond the resonance radius successfully develop both MRI turbulence and eccentricity.  Hence the Lubow mechanism is capable of exciting strong disk eccentricity even in the presence of strong MRI turbulence.  Eccentricity evolution results in considerable variability of the disk, and we elucidate the mechanisms of this variability.

This paper is organized as follows.  In \autoref{sec:method} we describe the overall setup, parameters, initial conditions, and boundary conditions of our simulations.  We then describe the evolution observed in our simulations in \autoref{sec:results}, both those that are initialized inside the resonance radius and those that are initialized beyond this radius.  In \autoref{sec:discussion}, we discuss the implications of our results, as well as some important caveats, and we summarize our conclusions in \autoref{sec:conclusions}.

\section{Method}
\label{sec:method}

Our 3D MHD simulations were conducted using cylindrical polar coordinates $(r,\phi,z)$ in the code
\textsc{Athena++}
\citep{athena}.
\textsc{Athena++} is designed to conserve the \(z\)-component of angular momentum in these coordinates, preserving numerical accuracy over long timescales.
To study the conditions experienced near
the disk midplane and simplify the problem, we chose our simulation domain to be a cylinder with vertically periodic
boundary conditions,
and neglected the vertical component of gravity. Inflow and outflow boundary conditions were chosen for the inner and outer radial boundaries, respectively.  The simulations employed
an accelerating, rotating reference frame in which the white dwarf was fixed at the origin and the binary companion along with its tidal potential remained static in time.
The numerical scheme by which the simulations were carried out was nearly identical to that in \citet{Oyang2021}, with the notable difference that we removed vertical stratification and used cylindrical polar coordinates.

The various parameters of the binary system are the same as those used in \citet{Oyang2021}. In particular, the white dwarf mass is 1.1~M$_\odot$, and the companion mass is 0.11~M$_\odot$.  The white dwarf radius is $4.69\times10^8$~cm, which we take to be our distance unit in the simulations.  The binary separation, distance to the L1 Lagrange point, and distance to the 3:1 orbital mean motion resonance are 32.68, 23.45, and 15.2 white dwarf radii, respectively.  The inner and outer cylindrical boundaries of the simulation are at the white dwarf radius and the radius of the L1 point, respectively.  In white dwarf radii the cylindrical coordinate grid spanned \(\left(r, \phi, z\right) \in \left[1, 23.44\right] \times \left[0, 2\pi\right) \times \left[-0.5, 0.5\right]\).  The number of cells in each direction was $N_r=384$ (logarithmically spaced), $N_\phi=768$, and $N_z=48$.  Time in all figures in this paper is measured in units of the binary orbital period (an ``orbit"), which is 938.5~s.

As already mentioned, the inner and outer radial boundaries have inflow and outflow boundary conditions, respectively, meaning density, azimuthal velocity, and pressure are copied from the active cells at the edges of the simulation domain into the ghost cells just across the boundaries. To prevent unwanted inflow of mass from the boundaries, the radial velocity is copied into the ghost cells only if it is flowing out of the simulation domain, otherwise the radial velocity in the ghost cells is set to zero. For the cases without an injected stream, magnetic fields in the ghost cells are also copied from the last active cells. When we inject a stream, we fix the magnetic field to be the initial condition in the region where the stream is injected, while magnetic fields in all the other ghost cells are still copied from the last active cell. 

We discuss five simulations in this paper, and some of their parameters are summarized in Table~\ref{tab:simparameters}.\footnote{Visualizations of the evolution of all simulations can be viewed at \href{https://www.youtube.com/playlist?list=PLKSZL0go-Er_JKCRTJ4CarJA2JMqxOh6M}{www.youtube.com/playlist?list=PLKSZL0go-Er\_JKCRTJ4CarJA2JMqxOh6M}.}
The first simulation, Hydro-$\alpha$,
was an alpha disk initialized with an empty simulation domain and then fed material with a stream from the Lagrange point on the outer boundary. It then evolved under the influence of hydrodynamic alpha viscosity and served as a useful point of comparison for the MHD simulations. This simulation was locally isothermal, which means the temperature was fixed in time and varied as $1/r$ so that the orbital Mach number $\mathcal{M}$ is fixed to be 20.

Simulation MHD-stream
was also initialized with an empty simulation domain and then was fed with a stream of mass and magnetic field from the Lagrange point on the outer boundary.  The stream had a Gaussian density profile with a width set to 0.5 white dwarf radii. Pure vertical magnetic fields were initialized inside the stream in order to seed the MRI. This magnetic field configuration is easy to achieve in cylindrical coordinates and differs from the magnetic field loops used in the MHD simulation of \citet{Oyang2021}. This simulation was also locally isothermal with orbital Mach number $40$.

Our remaining three simulations MHD-$\beta4$, MHD-$\beta3$, and MHD-$\beta1$
had no accretion stream, 
and were initialized with a disk with radial density profile $\rho\propto r^{-0.5}$ out to $r=16.53$. Beyond that radius, the density was exponentially dropped to the floor value with a characteristic length scale of $0.1$ white dwarf radii. At each radius, the initial density was uniform vertically. 
These simulations have orbital Mach number fixed to be 40, 20, and 20, respectively.
The magnetic field was initialized as a $5<r<16.33$ annulus of vertical field, with uniform plasma beta (See Table~\ref{tab:simparameters}).

\section{Results}
\label{sec:results}

\begin{figure}
    \includegraphics[width=6cm]{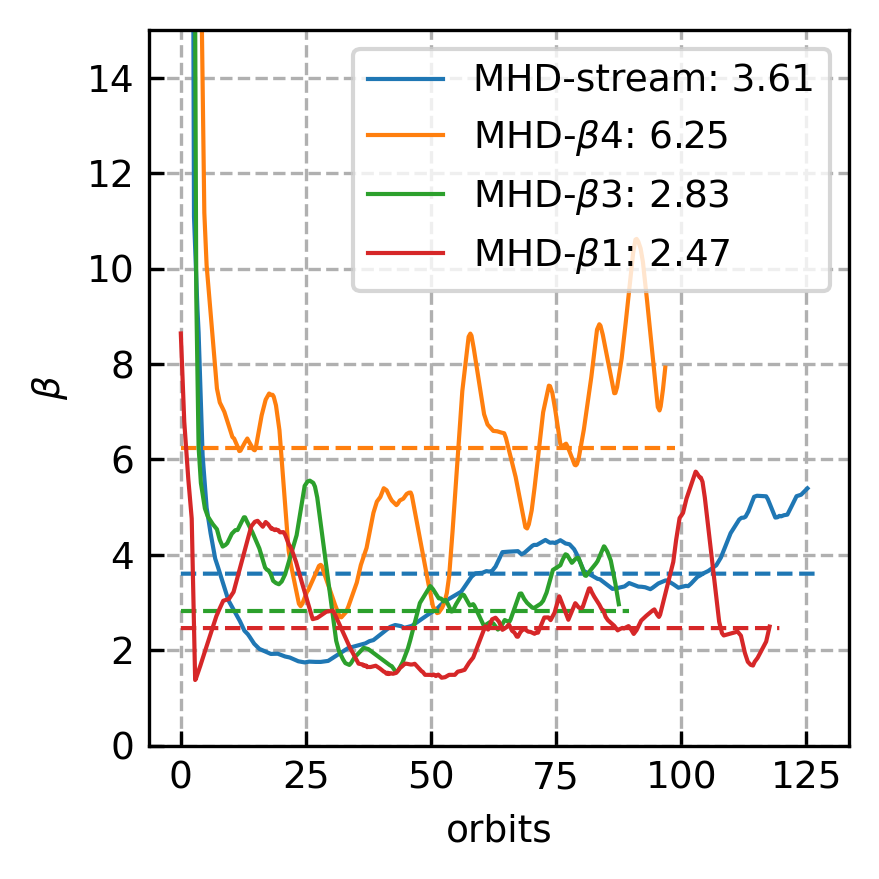}
    \includegraphics[width=12.4cm]{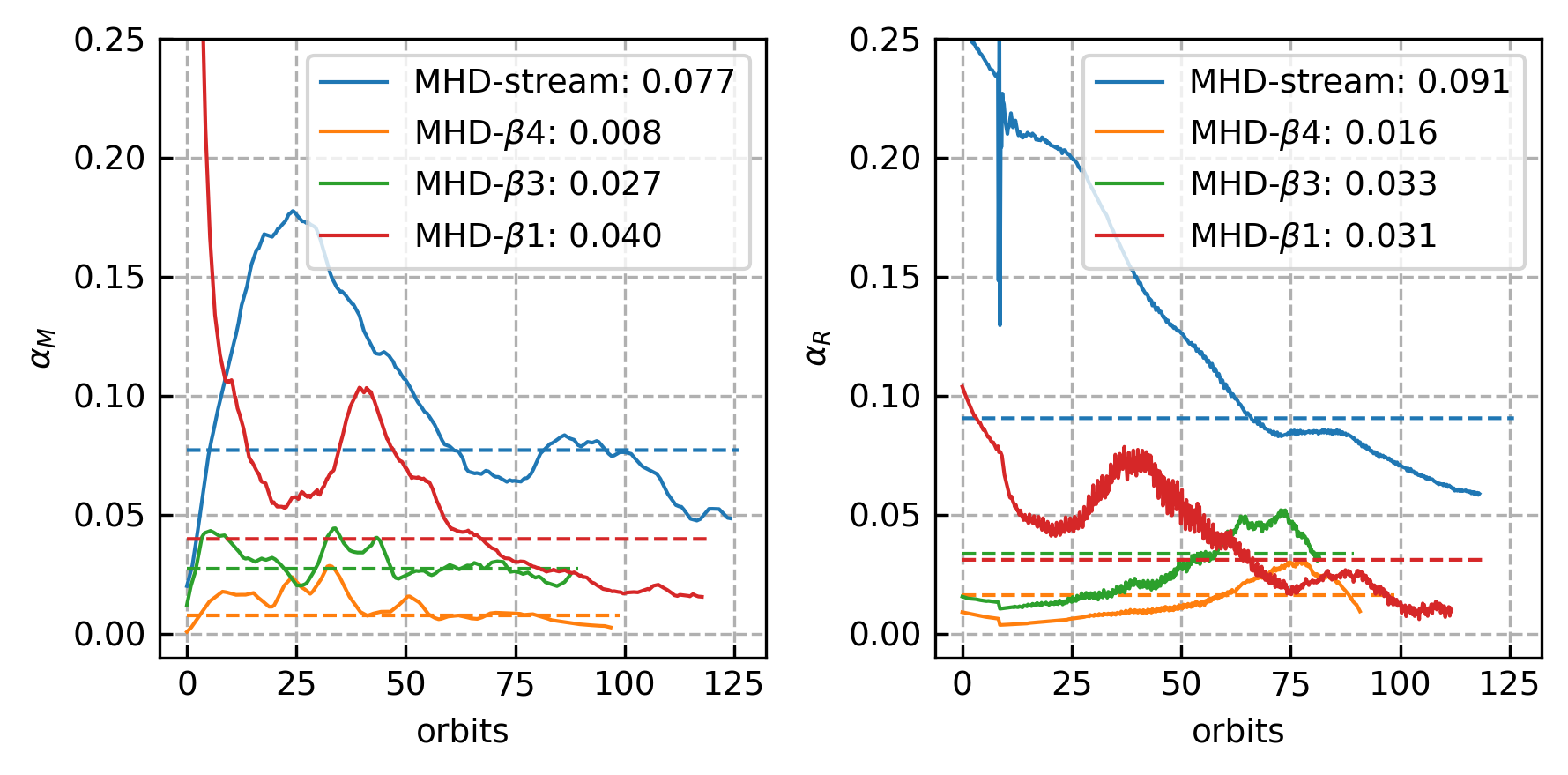}
    \caption{Time evolution of the spatially averaged plasma beta (left), Maxwell stress parameter (middle), and Reynolds stress parameter (right)
    in each of the four simulations. Horizontal dashed lines indicate time-averages after 35 binary orbital periods. The full evolution of all simulations can be viewed \href{https://www.youtube.com/playlist?list=PLKSZL0go-Er_JKCRTJ4CarJA2JMqxOh6M}{here}.}
    \label{fig:parameter_evolution}
\end{figure}

\autoref{fig:parameter_evolution} depicts the time-evolution of the spatial averages of the plasma beta parameter and dimensionless stress parameters in the four MHD simulations.
These spatial averages were computed over the entire simulation domain as follows:
\begin{eqnarray}
    &&\beta\equiv\frac{\int P_{\rm gas} r dr d\phi dz}{\int P_{B} r dr d\phi dz}\cr
    &&\alpha = \frac{\int \sigma_{r\phi} r dr d\phi dz}{\int P_{gas} r dr d\phi dz}
\end{eqnarray}
Here $P_{\rm gas}$ is the gas pressure, $P_B$ is the magnetic pressure, $\rho$ is the density, and $\sigma_{r\phi}$ is the $r\phi$ component of the Maxwell or turbulent Reynolds
stress.  We box-car time-average these spatial averages over four binary orbits to smooth out smaller time scale fluctuations before plotting them in \autoref{fig:parameter_evolution} in order to make the long term trends more clear.  
The turbulent Reynolds stress is computed in the usual way by subtracting off the mass-weighted mean radial and azimuthal velocities before computing the average of the density times their product.  It is not clear that this is optimal if the disk has become eccentric, and \citet{chan22,chan23} avoided calculating the turbulent Reynolds stress for this very reason.

Table~\ref{tab:simparameters} lists the initial and time-averaged evolved plasma beta parameters, as well as the evolved stress parameters, in all four of our MHD simulations. All time averages were done only with data after 35 binary orbital periods, after transients associated with the initial conditions have died down.  As would be expected, in the simulations without an accretion stream (MHD-$\beta4$, MHD-$\beta3$, and MHD-$\beta1$)
the Maxwell alpha stress parameter is larger, and the time-averaged plasma beta is smaller, for smaller initial plasma beta in the initial vertical flux.  In other words, stronger initial magnetic flux leads to stronger magnetic pressure and stresses in the MRI turbulent state.

\begin{table}
	\centering
	\caption{Simulation parameters.  The symbol $<>$ denotes averages of the plasma $\beta$, and the $r\phi$ Maxwell and turbulent Reynolds stress components,
    expressed as Shakura-Sunyaev alpha parameters $\alpha_M$, and $\alpha_R$,
    respectively.  These averages are done over the entire spatial domain and over time after 35 binary orbital periods.  Values before 35 binary orbital periods are also ignored while computing the maximum eccentricity.}
	\begin{tabular}{clcccc}
		\hline
		Name 
            & Hydro-$\alpha$ 
            & MHD-stream 
            & MHD-\(\beta\)4 
            & MHD-\(\beta\)3 
            & MHD-\(\beta\)1 
            \\
            \hline
              $\mathcal{M}$ & 20 & 40 & 40  & 20 & 20 \\
            \hline
            Initial \(\beta\) & N/A & N/A & \(10^4\) & \(10^3\) & 10 \\
		\hline
		  $\langle\beta\rangle$ & N/A & 3.61 & 6.25 & 2.83 & 2.47 \\
		\hline
		  $\langle\alpha_M\rangle$ & N/A & 0.077 & 0.008 & 0.027 & 0.040 \\
            \hline
        		  $\langle\alpha_R\rangle$ & N/A & 0.091 & 0.016 & 0.033 & 0.031 \\
		\hline
		\hline
            $\alpha$ & 0.1 & N/A & N/A & N/A & N/A \\
		\hline
            \(\max\langle e \rangle\) & 0.109 & 0.010 & 0.137 & 0.164 & 0.203 \\
		\hline
	\end{tabular}
    \label{tab:simparameters}
 \end{table}

\begin{figure*}
    \includegraphics[width=9cm]{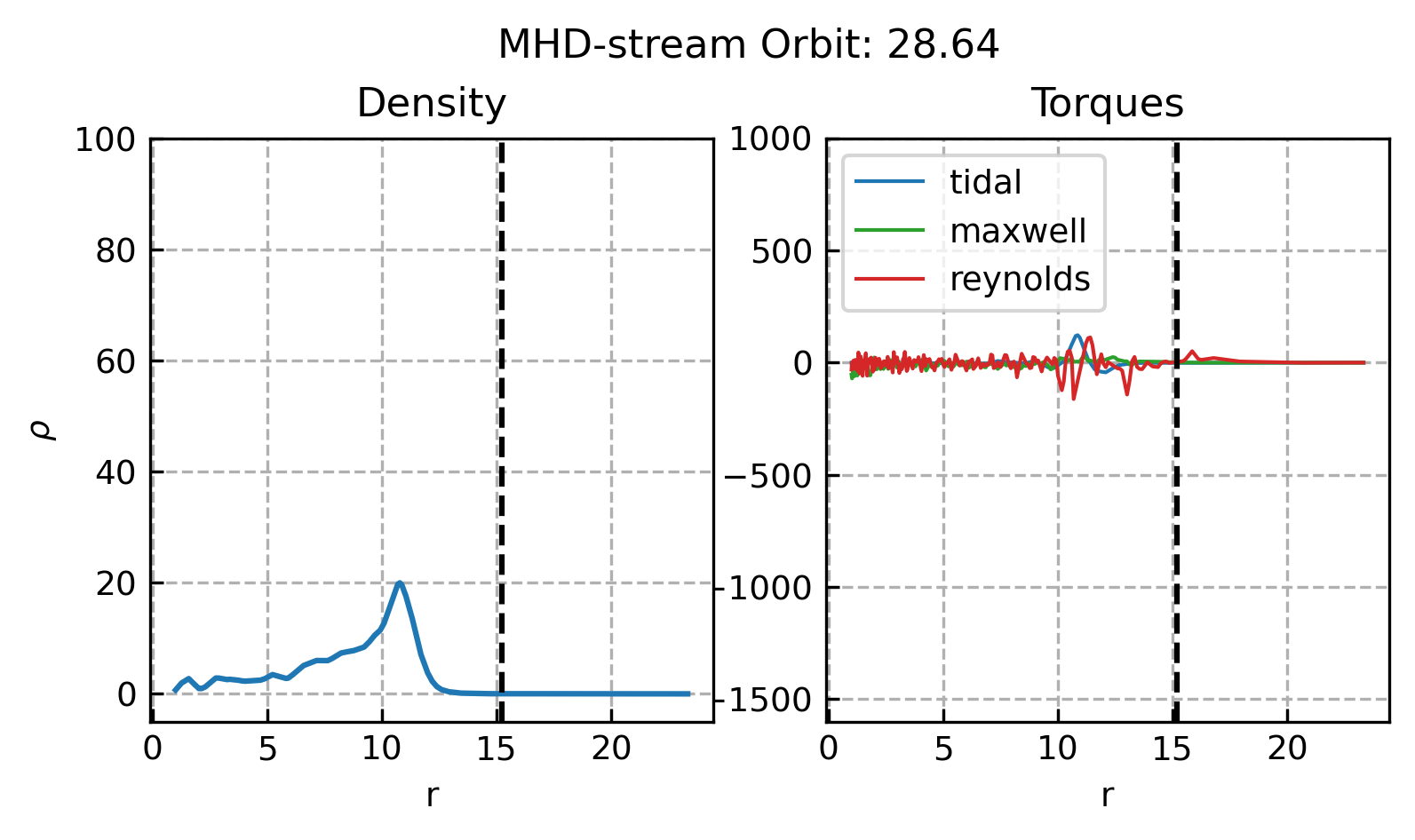}
    \includegraphics[width=9cm]{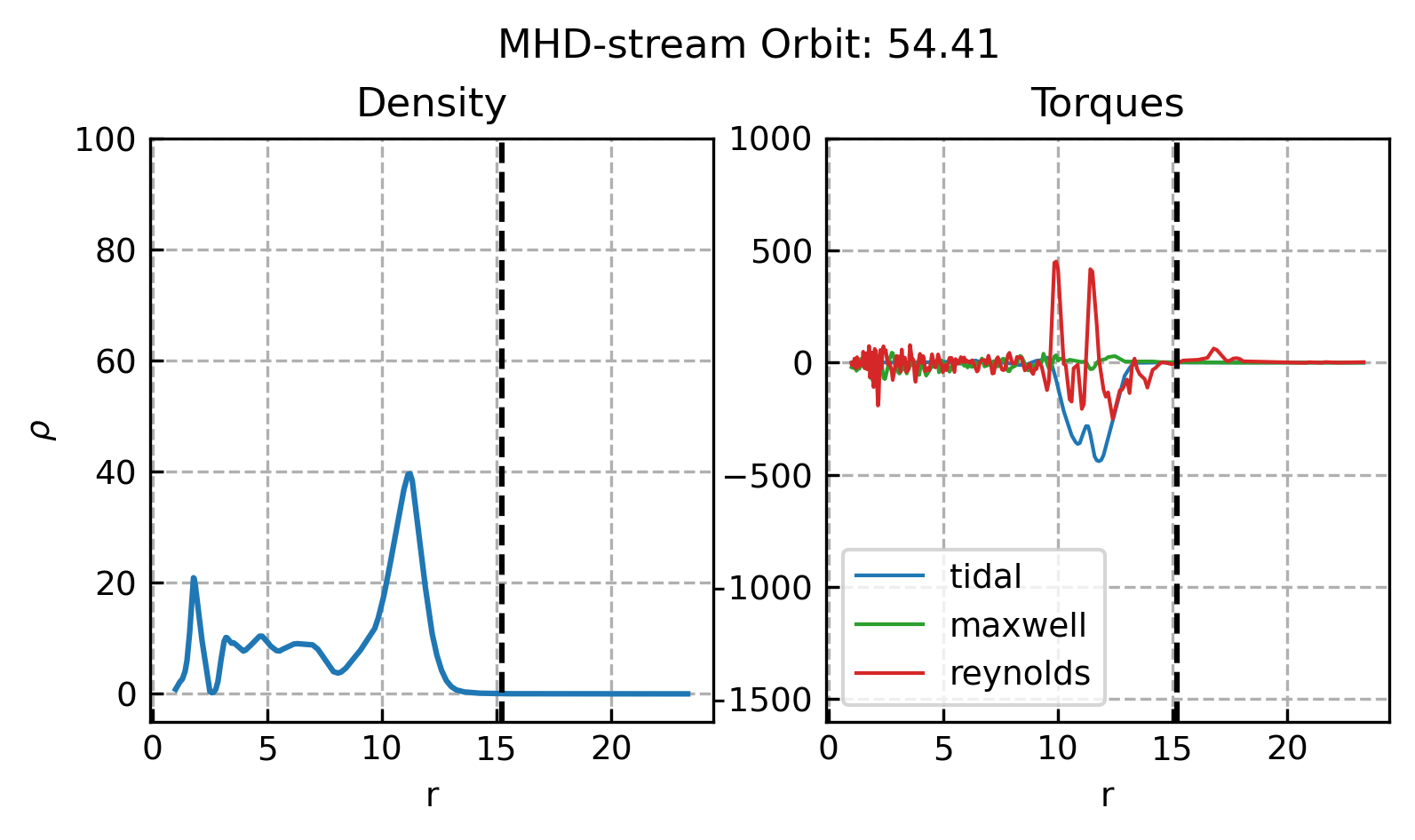}
    \begin{center}
    \includegraphics[width=10cm]{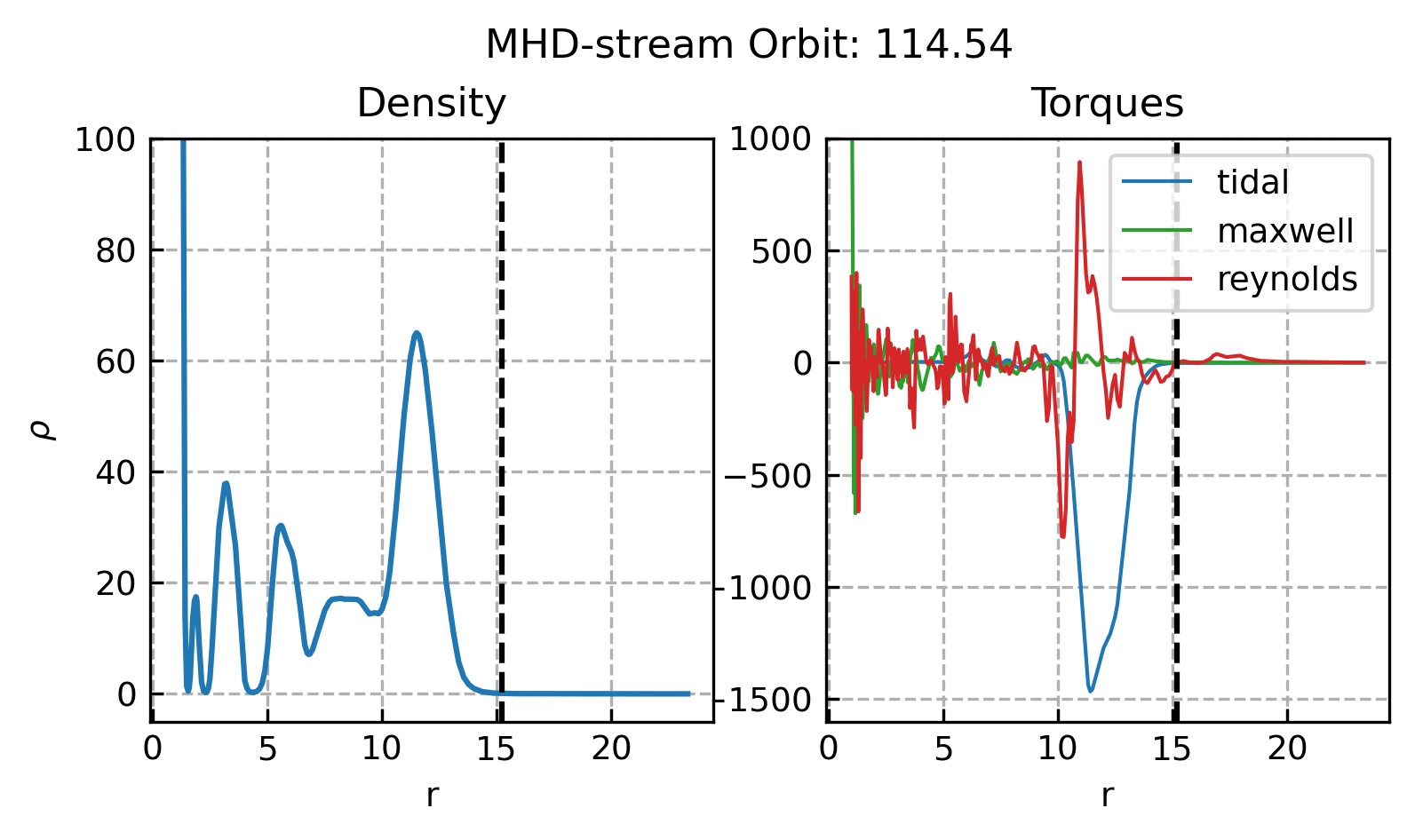}    
    \end{center}
    \caption{Radial dependence of vertically and azimuthally-averaged density (left panel) and tidal and turbulent torques (right panel) in MHD-stream 
    at 28.64 (upper left), 54.41 (upper right), and 114.54 (bottom) binary orbital periods after the start of the simulation. Outward radial spreading of the disk stalled, due to the overwhelming tidal torques on the density spike in the outer disk. Because the disk was unable to reach the resonance radius (dashed line), no significant eccentricity developed.
    }
    \label{fig:cyl_2_torques}
\end{figure*}

Simulation MHD-stream
builds up the disk with an accretion stream in a similar manner to \citet{Oyang2021}.  With no vertical stratification, however, we were able to inject a net poloidal magnetic flux that resulted in enhanced MRI torques: the average total (Maxwell plus Reynolds)
alpha parameter was 0.168, compared to a total alpha parameter of $\simeq0.01$ in the MHD simulation of \citealt{Oyang2021}.  Nevertheless, this simulation still failed to develop significant eccentricity.  As we will show later in disks with MRI turbulence that do grow eccentricity, this was not because
MHD may have been inhibiting the Lubow mechanism.
Instead, MHD disks are simply harder to spread than alpha disks. The reason for this is that the tidal torques which act toward truncating the disk are naturally proportional to the density in the disk, and in an alpha disk model the viscous torques are proportional to pressure, and thus density as well. (Radiation pressure, which does not depend on density, is never relevant to the outer disk in compact binary systems.)  This means that viscous and tidal torques scale together with density so the disk structure can largely be ignored when studying how an alpha disk spreads. In contrast, however, the Maxwell stress torques are not proportional to density but instead to field strength which is only indirectly related to density. As such, the disk structure cannot be neglected. Specifically if an overdense ring forms, e.g. due to stalling of the outward spreading of the disk, tidal torques will strongly dominate Maxwell stress torques and truncate the disk early.  That this happened in MHD-stream
can be clearly seen in Figure \ref{fig:cyl_2_torques}.   Note that turbulent Reynolds stresses will still naturally be proportional to density, and Figure~\ref{fig:cyl_2_torques} shows that they do grow with time in MHD-stream in the outer, stalled, dense ring.  But they still fail to overcome the tidal torque.  We believe this to be the fundamental problem in MHD-stream
and the MHD simulation of \citet{Oyang2021}:  MHD does not inherently suppress eccentricity; it just makes spreading the disk to the resonance radius a more complex endeavour.

\begin{figure*}
    \includegraphics[width=9cm]{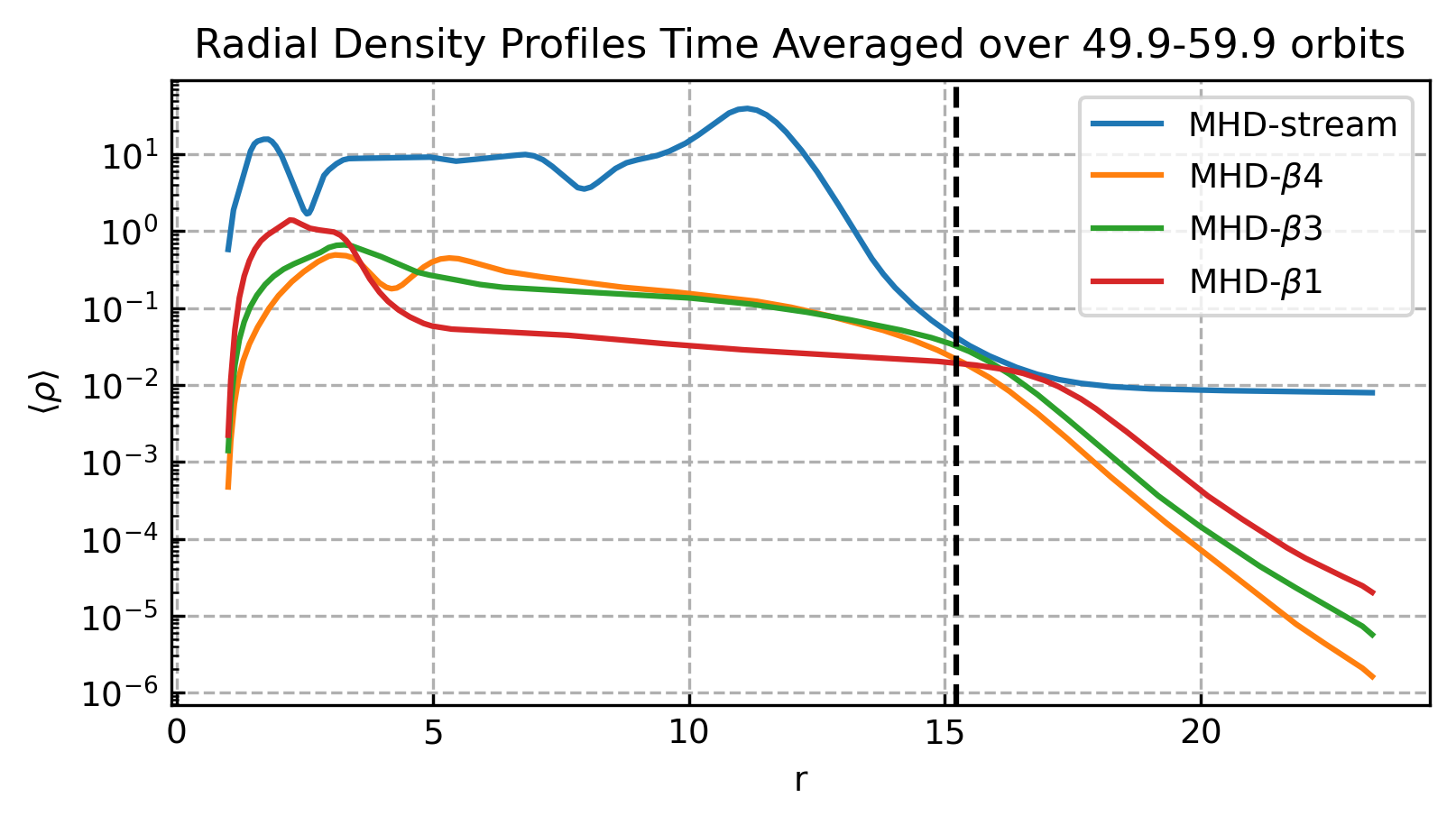}
    \includegraphics[width=9.5cm]{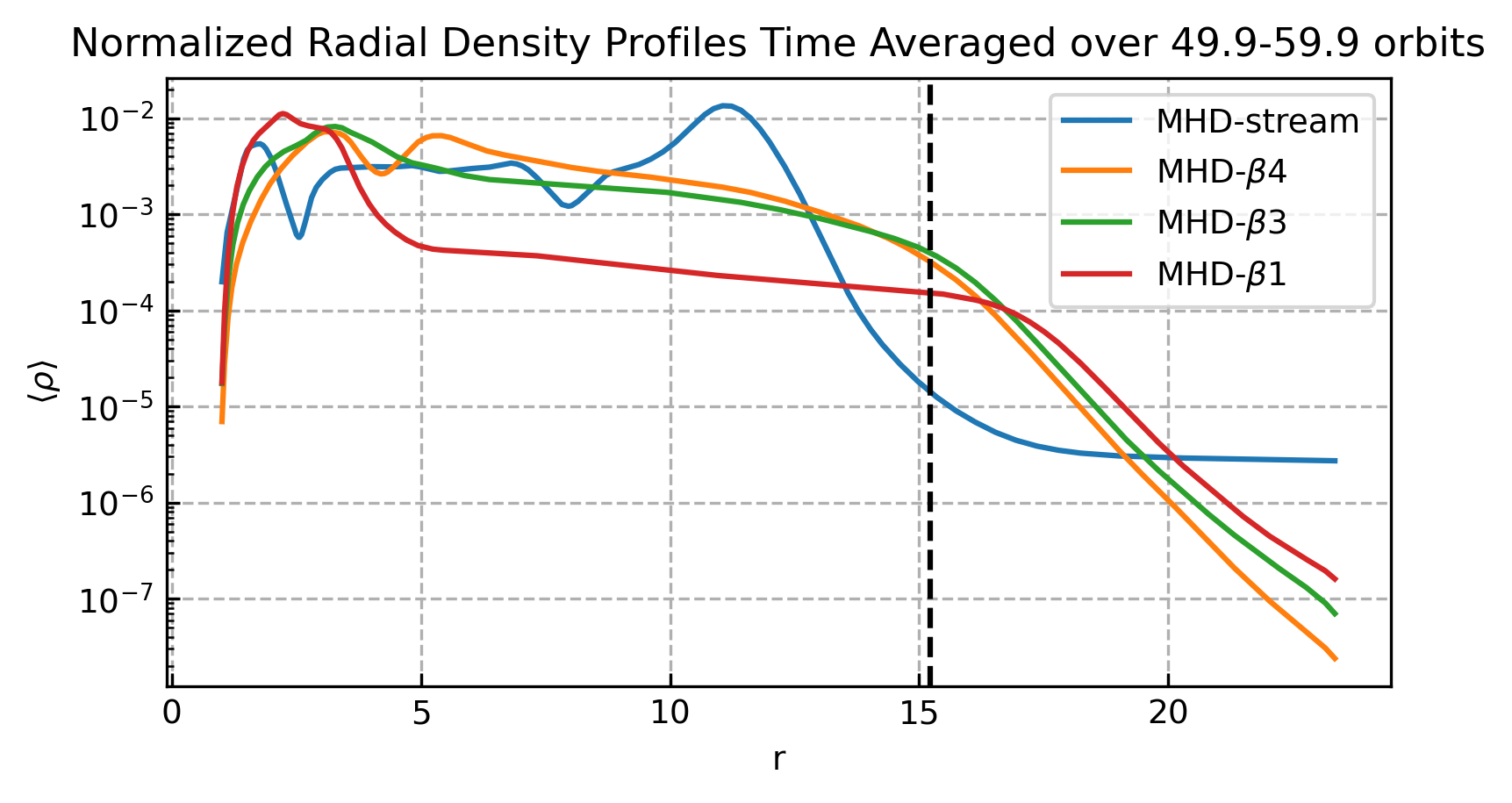}
    \caption{Radial profiles of vertically and azimuthally-averaged density, time-averaged over ten (49.9-59.9) binary orbits.  The left panel shows the actual density profiles, while the right panel renormalizes the density so that the disks all have the same total mass.  The resonance radius is indicated by the vertical dashed line.
    }
    \label{fig:radial_density_profiles}
\end{figure*}

Having concluded that spreading the disk was the fundamental issue with lack of eccentricity growth, we conducted simulations which were initialized at the resonance radius to directly see how the MRI interacted with the Lubow mechanism. To make the numerical experiment as clean as possible, we also removed the accretion stream in these simulations, as constant feeding by a stream tends to damp eccentricity \citep{lubow_eccentricity_damping_by_stream,kle08}.  In all these simulations we saw that MRI was able to maintain the disks out at the resonance radius and allow eccentricity to develop exactly as described by the Lubow mechanism.  \autoref{fig:radial_density_profiles} compares the radial density profiles between approximately 50 and 60 binary orbits across all the MHD simulations.  While all the simulations have approximately the same density at the resonant radius, MHD-stream, which by this time has much more mass, has a more strongly radially declining profile than the others.  This apparently inhibits waves excited at the resonance from propagating inward into the body of the disk.

Note that even simulation MHD-$\beta$4, which had a time-averaged total alpha stress parameter of only 0.024, is able to maintain itself out at the resonant radius.  This is still larger than the alpha value  $\simeq0.01$ of the stream-fed MHD simulation in \citet{Oyang2021}, which failed to spread out to the resonance radius and grow eccentricity.  It appears that even relatively weak Maxwell stresses (as small as 0.024) are able to keep the disk out at the resonant radius and grow eccentricity, provided the disk was initialized there.  Spreading the disk out to the resonant radius is the more serious challenge.

\begin{figure*}
    \includegraphics[width=9cm]{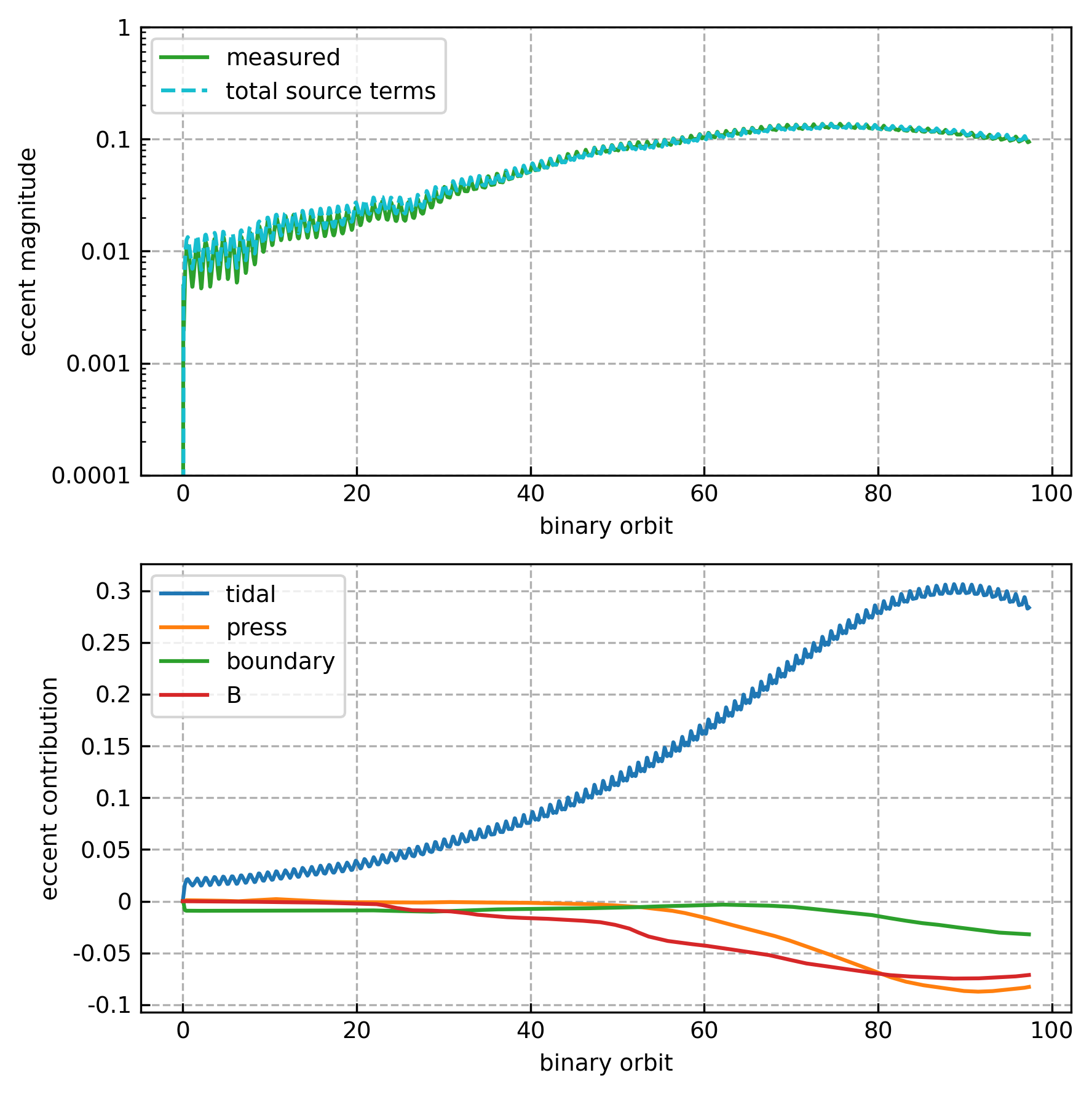}
    \includegraphics[width=9cm]{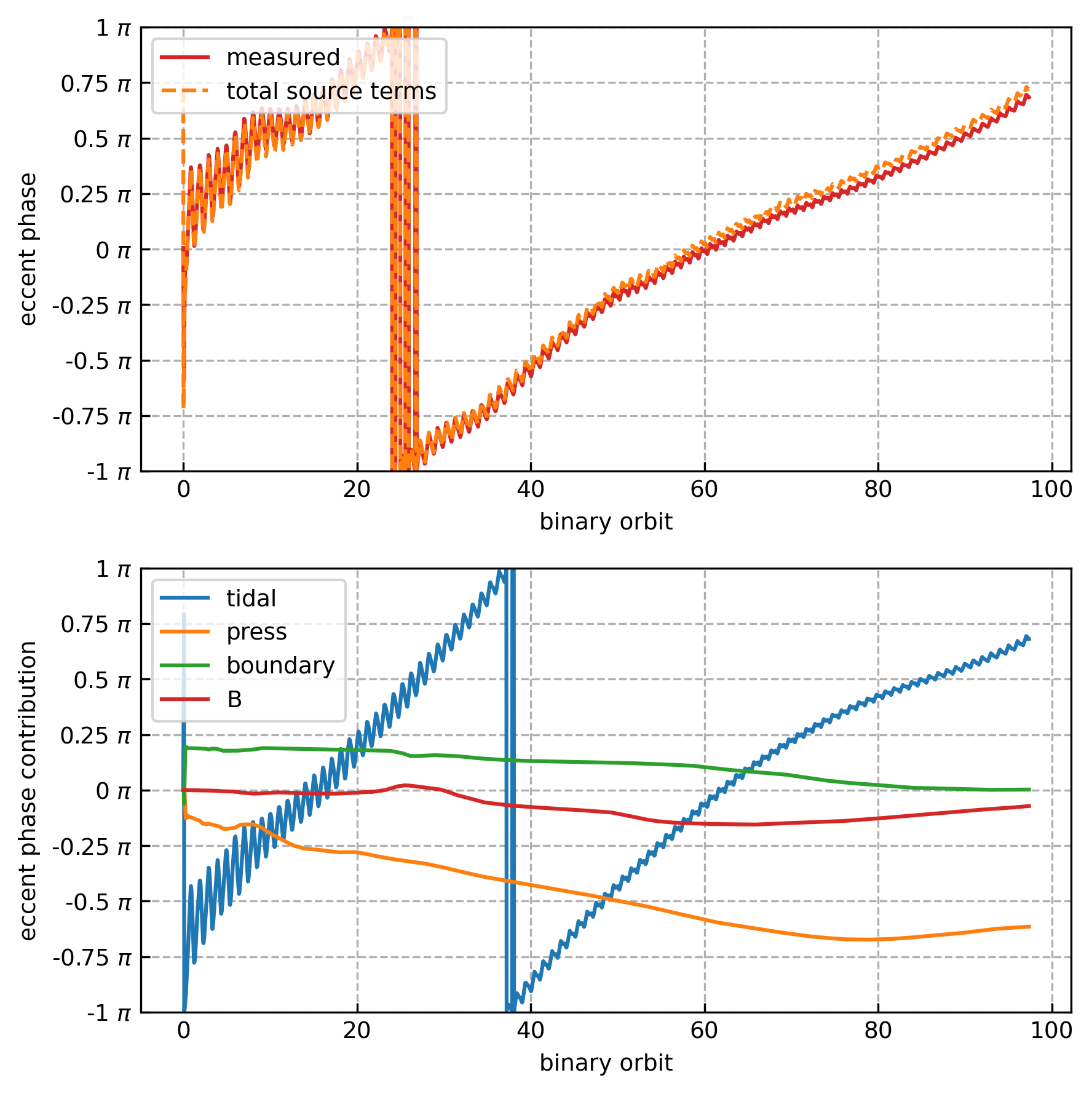}
    \caption{Evolution of eccentricity for simulation MHD-$\beta4$.
    The left and right columns of panels refer to the eccentricity magnitude and direction (in terms of the Laplace-Runge-Lenz vector, measured in a non-rotating frame with origin $r=0$).  The top panels show the overall evolution of these quantities, and the bottom panels show the cumulative contributions to this evolution from tidal forces, pressure forces, magnetic forces, and losses through the boundaries.  The sum of these source terms is also shown in the top panels, and is in good agreement with the total measured evolution.}
    \label{fig:Cyl_11_2_growth_prec}
\end{figure*}

\autoref{fig:Cyl_11_2_growth_prec} shows the global eccentricity evolution in simulation MHD-$\beta4$.
This is based on the following evolution equation
\begin{equation}
    \partial_t(\rho\boldsymbol{e})+\boldsymbol{\nabla}\cdot({\bf v}\rho\boldsymbol{e})=\frac{1}{GM_1}\left[\boldsymbol{f}\times(\boldsymbol{r}\times\boldsymbol{v})+\boldsymbol{v}\times(\boldsymbol{r}\times\boldsymbol{f})\right],
\end{equation}
where $\boldsymbol{f}$ is the force per unit volume summed over all forces acting on fluid elements, apart from the $1/r^2$ force from the primary mass $M_1$ at the origin which does not source eccentricity \citep{Oyang2021}.  The quantity $\boldsymbol{e}$ is the eccentricity vector (proportional to the Laplace-Runge-Lenz vector), given by
\begin{equation}
    \boldsymbol{e}=\frac{1}{GM_1}\boldsymbol{v}\times(\boldsymbol{r}\times\boldsymbol{v})-\hat{\boldsymbol{r}}.
\end{equation}
The global eccentricity and phase are based on the magnitude and direction of the mass-weighted volume average of this eccentricity vector:
\begin{equation}
    \langle\boldsymbol{e}\rangle=\frac{\int_V\rho\boldsymbol{e}dV}{\int_V\rho dV}
\end{equation}

The tides successfully grow eccentricity in the disk despite the fact that magnetic and pressure forces act as eccentricity sinks, as shown previously in \citet{Oyang2021}.  That MRI turbulence itself would act as an eccentricity sink is to be expected since any dissipative effect will tend to circularize orbits, but this effect is small compared to the tidal driving, and eccentricity grows in this simulation. As we will discuss in more detail below (see \autoref{fig:eccent_comp}), simulations MHD-$\beta$3 and MHD-$\beta$1, which have even larger Maxwell stress alpha parameters, also grow eccentricity in a similar fashion.  As the tides build up eccentricity, the system eventually saturates at a maximum eccentricity a bit above 0.1 after which the precession slows down. Additionally, in contrast to the alpha disk, the MHD turbulence seems to favor build up of eccentricity in the inner regions of the disk at the expense of eccentricity in the outer regions of the disk. This is exactly consistent with the results of \citep{chan22}, as shown in their figure 4.

\begin{figure*}
    \includegraphics[width=9cm]{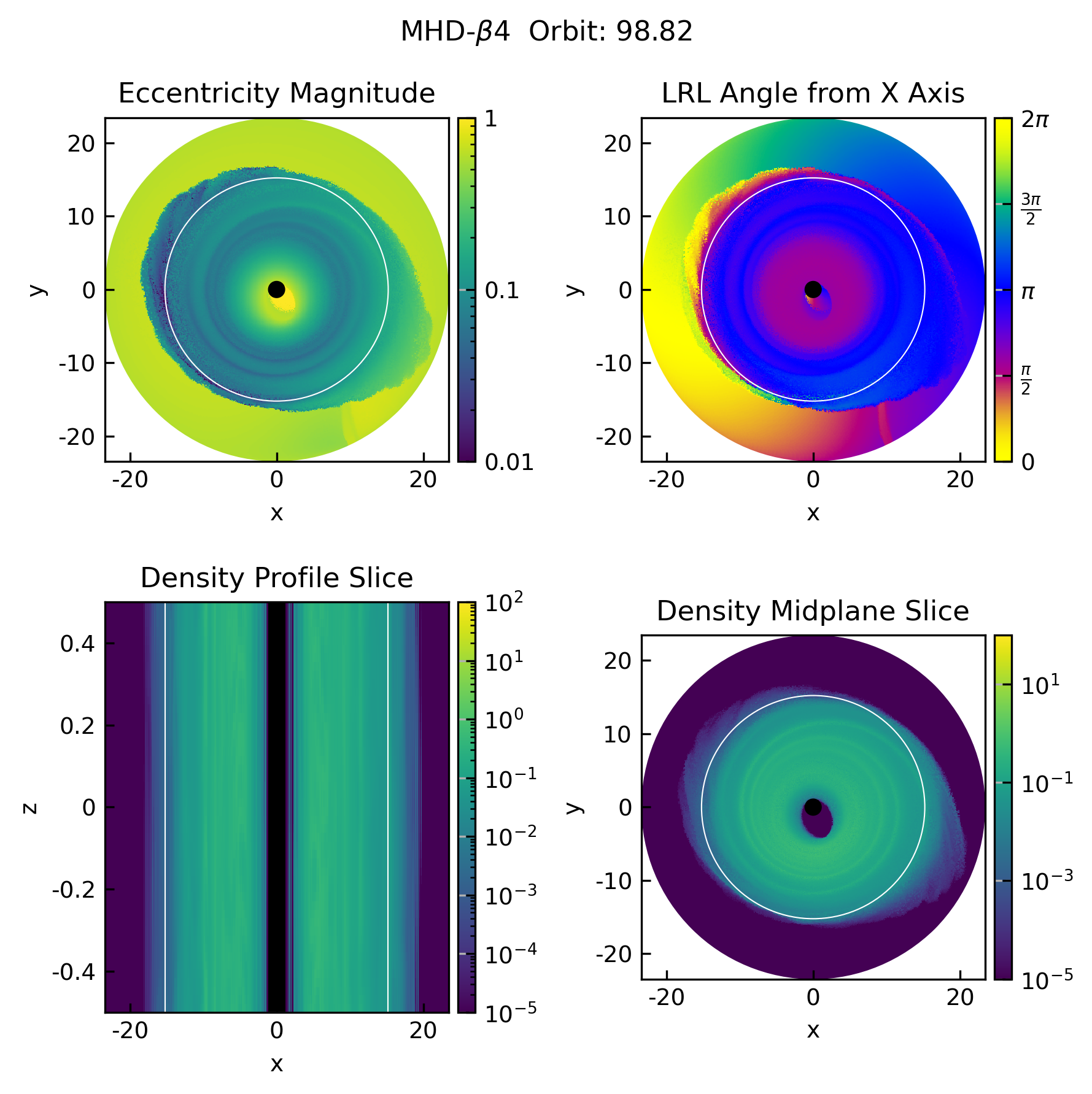}
    \includegraphics[width=9cm]{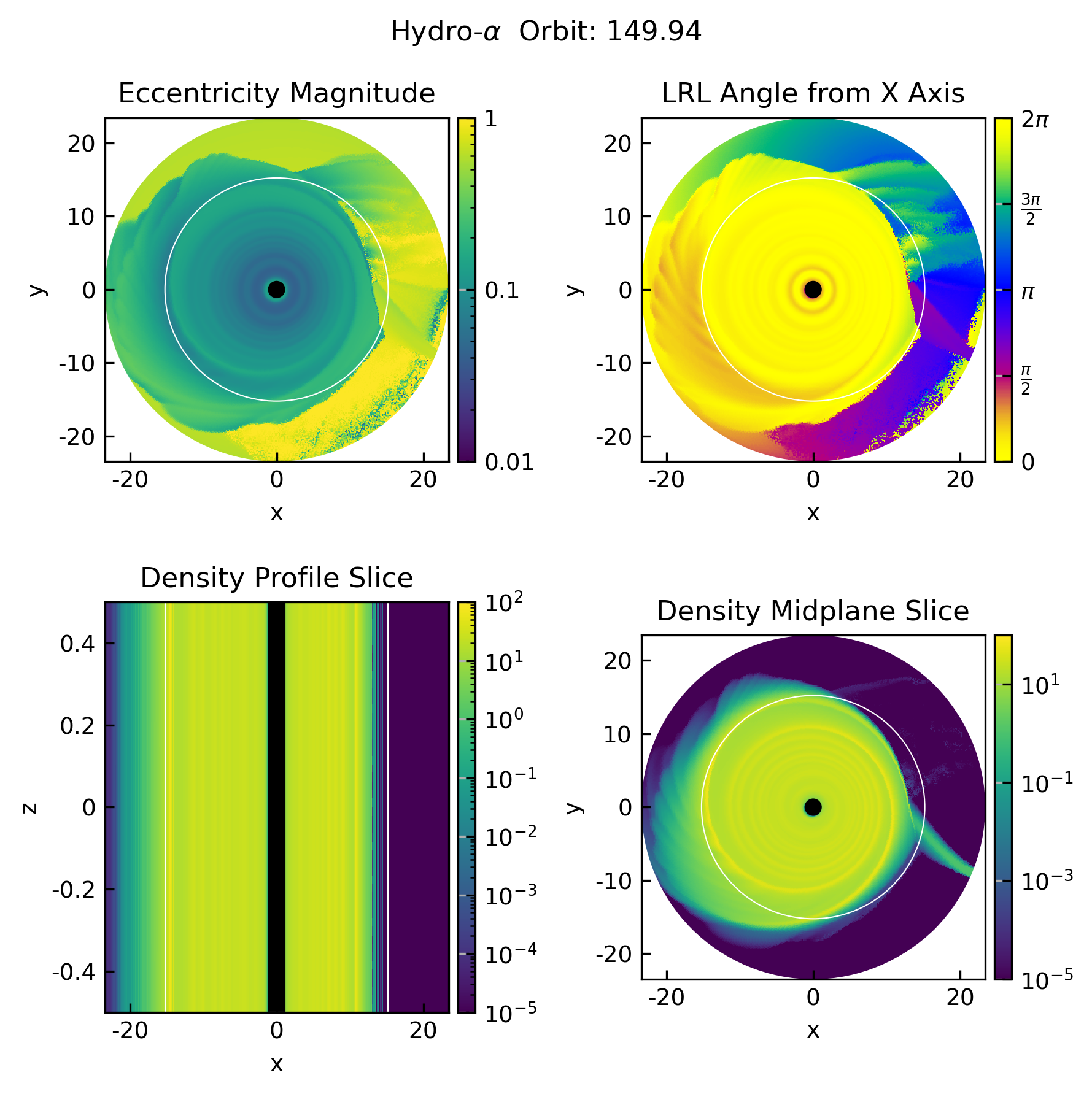}
    \caption{Late time evolution of an MHD simulation (MHD-$\beta4$,
    left group of four panels) and the stream-fed alpha disk (Hydro-$\alpha$,
    right group of panels).  The 3:1 mean motion resonance is indicated by the white lines in all the figure panels.  Going clockwise from the upper left in each group, the panels show the magnitude of the local eccentricity in the $z=0$ plane, the angle between periapsis and the $x$-axis in the $z=0$ plane, the density in the $z=0$ plane, and the radial and vertical distribution of density in the $y=0$ plane.}
    \label{fig:late_time_1_and_11}
\end{figure*}

\autoref{fig:late_time_1_and_11} compares the two-dimensional density and eccentricity distributions in the $z=0$ plane at late times in the MHD simulation MHD-$\beta4$
with those in the hydrodynamic alpha-disk simulation Hydro-$\alpha$.
In a standard alpha disk under a tidal perturbation, inward propagating waves excited at the resonance facilitate the smooth growth of eccentricity throughout the disk.  In contrast, in the MHD case we see two distinct inner and outer eccentric regions which are misaligned, and separated by a region of circularity.  The misalignment of the two regions in MHD-$\beta4$
is most apparent in the orientation of the Laplace-Runge-Lenz vector, whereas that vector is aligned everywhere in the disk in Hydro-$\alpha$.
We suspect that the boundary between the two misaligned regions in MHD-$\beta4$
is kept circularized by the dissipation resulting from orbit crossings between these two regions.  Note that an eccentric void has formed at the center of MHD-$\beta4$
due to loss of material through the inner boundary as the pericenters of the eccentric orbits reach that boundary.
In fact, all the MHD simulations that form eccentric disks also form eccentric inner voids, and these continue to grow even after eccentricity has saturated and stopped growing.  In fact, as we will discuss later, in MHD-$\beta1$ the void does not even form until after that disk has reached its maximum eccentricity.
No such void ever forms in the alpha-disk simulation Hydro-$\alpha$.

\citet{chan22} also observe inner eccentric voids in simulations of eccentric disks with and without magnetic fields.  It is important to note that their eccentric simulations were initialized with eccentricity, with an eccentric void in the middle.  In the case of their MHD simulations, the eccentricity of this void grew somewhat as material at periapsis was lost through the inner boundary, entirely consistent with what we see here.  However, the persistence of the voids in their eccentric hydrodynamic simulations contrasts with the fact that we do not observe voids in our simulation Hydro-$\alpha$.  They did not include an alpha-viscosity in their simulations, and it may be
that this would have caused their hydrodynamic disk to spread inward and fill its void.  It must of course be noted that both our simulations and those of \citet{chan22} have inflow boundary conditions at the inner boundary.  If the central mass is a star such as a white dwarf, a boundary layer would form between the disk and the star and would
likely have a strong effect on the formation and persistence of a void in Nature.   Indeed, \citet{kle08} also found that eccentric voids could form in their hydrodynamical viscous disk simulations (see their Figure 17), but only when an inflow inner boundary condition was used.

\begin{figure*}
    \includegraphics[width=16cm]{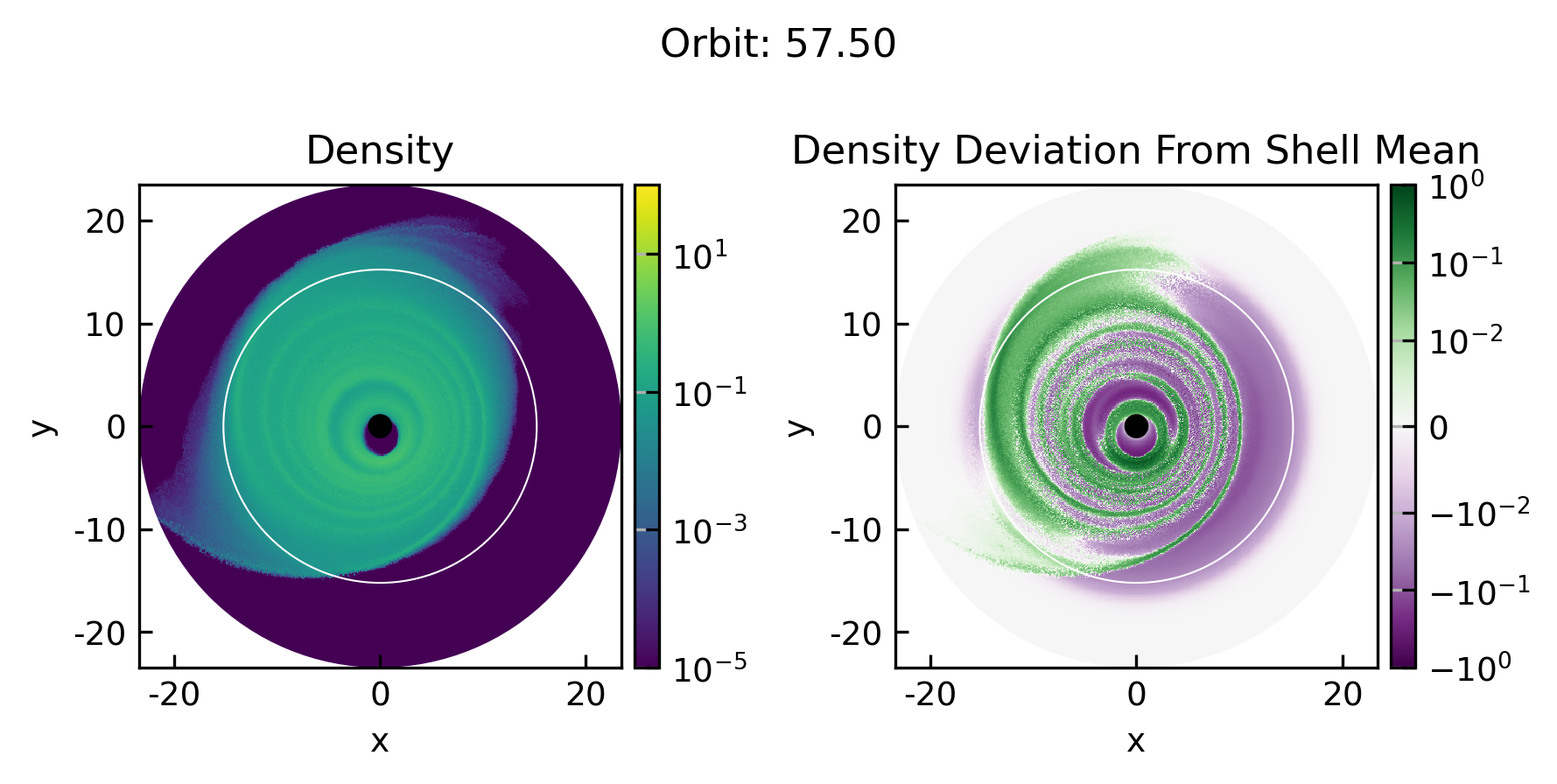}
    \centering
    \caption{Density in the $z=0$ midplane (left), and its deviation from the vertically- and azimuthally- (shell) averaged density (right) in MHD-$\beta4$.
    Spiral waves in the outer radial regions, and eccentric standing modes in the inner radial regions, are apparent in the right hand panel.}
    \label{fig:cyl_11_2_waves}
\end{figure*}

The misalignment between the inner and outer disks comes and goes with time, and the radius that separates them is not fixed.   \autoref{fig:cyl_11_2_waves} shows more detail of the density distribution in the MHD simulation MHD-$\beta4$
at a somewhat earlier time, when the radius between the misaligned inner and outer disks is smaller.  In the outer region we see the spiral waves driven by the tidal resonance as predicted by the Lubow mechanism. In the inner, misaligned region, we do not see spiral waves but instead elliptical standing wave modes reminiscent of those described in \citet{ogilvie_lynch_2018}. This suggests that eccentricity, which is being sourced by the resonance in the outer disk, is somehow being moved inward towards the inner disk where it is deposited into the standing waves of the eccentric core.

\begin{figure*}
    \includegraphics[width=18cm]{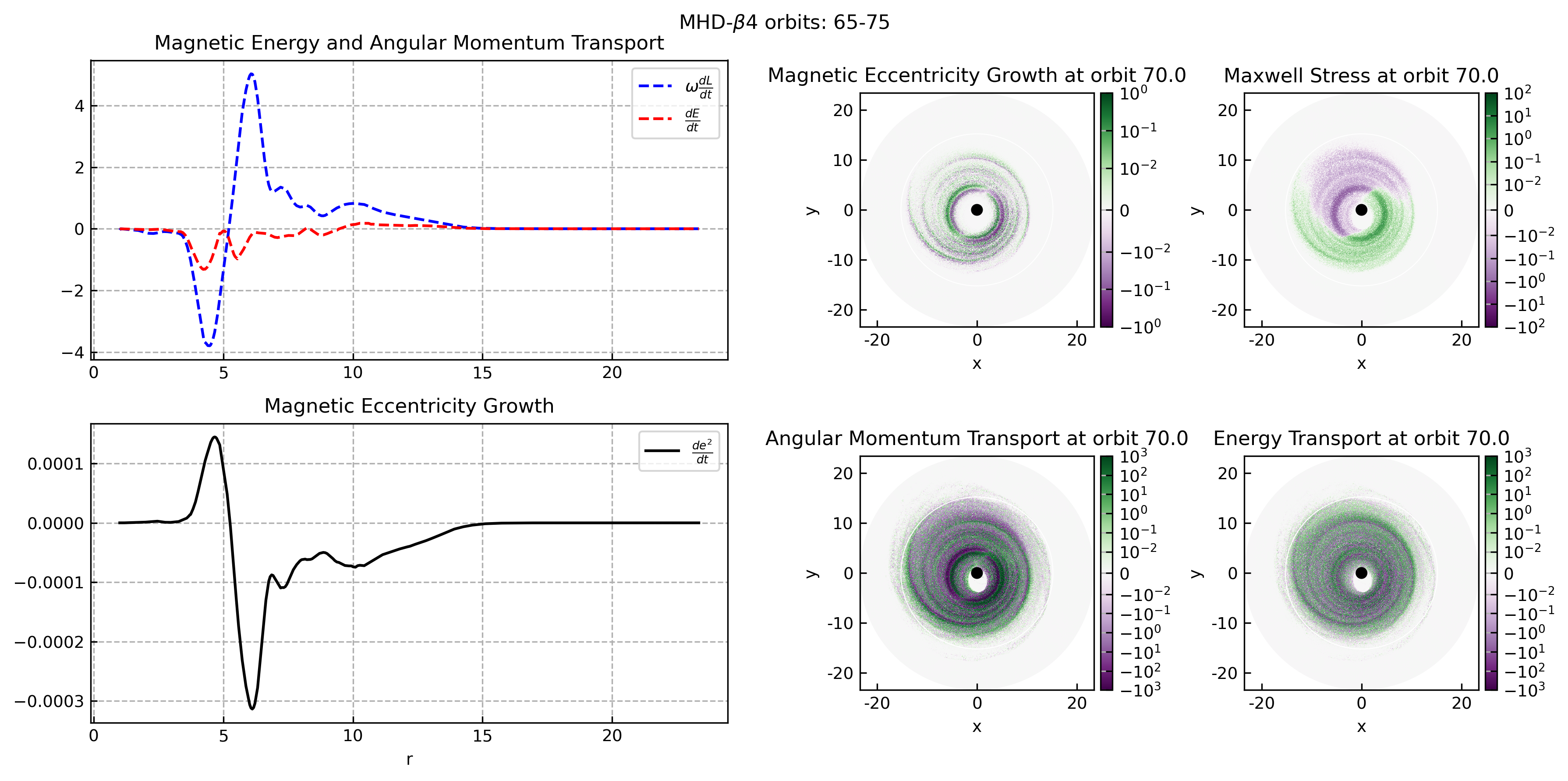}
    \caption{Left panels: local rate of work done by magnetic forces and the local Maxwell stress torque times local angular velocity (top), vertically and azimuthally-averaged as well as time-averaged over epochs 65-75 orbits in simulation MHD-$\beta$4.
    The predicted eccentricity growth from equation (\ref{eq:de2dt}) is shown in the bottom panel.  Right panels:  two-dimensional distributions of midplane slices of the eccentricity growth (upper left), Maxwell stress $-B_rB_\phi/(4\pi)$ (upper right), magnetic angular momentum transport $\omega dL/dt$ (lower left), and magnetic energy transport $dE/dt$ (lower right).}
    \label{fig:Cyl_11_2_flux}
\end{figure*}

That this is in fact happening can be seen in \autoref{fig:Cyl_11_2_flux}.  In circular disks, MRI turbulent stresses transport energy and angular momentum outward in such a manner as to maintain material on approximately circular orbits as it slowly spirals inward.  If $E$ and $L$ are the specific energy and angular momentum of a test particle orbit around a gravitating mass $M$, then the rate of change of the eccentricity of this orbit is given by
\begin{equation}
    \frac{de^2}{dt}=\frac{2L^2}{G^2M^2}\left(\frac{dE}{dt}-\frac{\omega}{\sqrt{1-e^2}}\frac{dL}{dt}\right)\simeq\frac{2L^2}{G^2M^2}\left(\frac{dE}{dt}-\omega\frac{dL}{dt}\right)
    \label{eq:de2dt}
\end{equation}
for small eccentricities, where $\omega$ is the mean angular velocity of the orbit.  In circular disks, MRI turbulent stresses maintain $dE/dt=\omega dL/dt$ so that eccentricity does not grow, but as pointed out by \citet{chan22}, this need not be the case in eccentric disks.  \autoref{fig:Cyl_11_2_flux} shows how differing rates of angular momentum and energy transfer by MRI turbulence lead to eccentricity build up in the inner regions.  The angular momentum term $\omega dL/dt$ in the upper left panel was computed from the Maxwell stress torque times the circular Keplerian angular velocity.  The energy term $dE/dt$ was computed from the scalar product of the local Maxwell force density times the local fluid velocity.  For circular orbits, these would cancel, but this is not true for eccentric orbits.  Note that the eccentricity growth shown in the bottom left panel is actually the maximum possible growth as predicted by equation (\ref{eq:de2dt}), as it assumes that the growth of eccentricity is coherent in terms of its orientation.  MRI turbulence could in principle excite eccentricities with random orientations which would then dissipate due to orbit collisions, but it appears that the average effect here produces a coherent orientation in the inner disk. Nevertheless, the upper left of the four right hand panels of \autoref{fig:Cyl_11_2_flux} shows that the magnetic stresses produce considerable azimuthal variation with both growth and decay of eccentricity.  Remarkably, in the innermost regions, the magnetic energy and angular momentum extraction (which individually do not vanish - see the lower two right hand panels) exactly cancel any eccentricity growth everywhere (upper left plot).  We suspect that this is due to the dominance of hydrodynamic eccentric waves.  An alternative would be that MRI stresses are suppressed by these waves \citep{dew20}, and there is some evidence for this in the reduced Maxwell stress in the upper right panel.  However, cancelation must be present to zero out the 2D magnetic eccentricity growth in the upper left panel.  In any case, near the outer edge of the inner disk, MRI stresses extract angular momentum from the eccentric orbits more effectively than energy, causing the eccentricity there to grow further.  At the same time, angular momentum is added to the outer disk, which would act to reduce eccentricity were it not for the fact that the Lubow mechanism resupplies it from the resonant radius.  Misalignment then results because of differential apsidal precession of the inner and outer disk.  Coherence in eccentricity orientation in the inner disk is likely maintained by the eccentric standing waves.

\begin{figure*}
    \includegraphics[width=12cm]{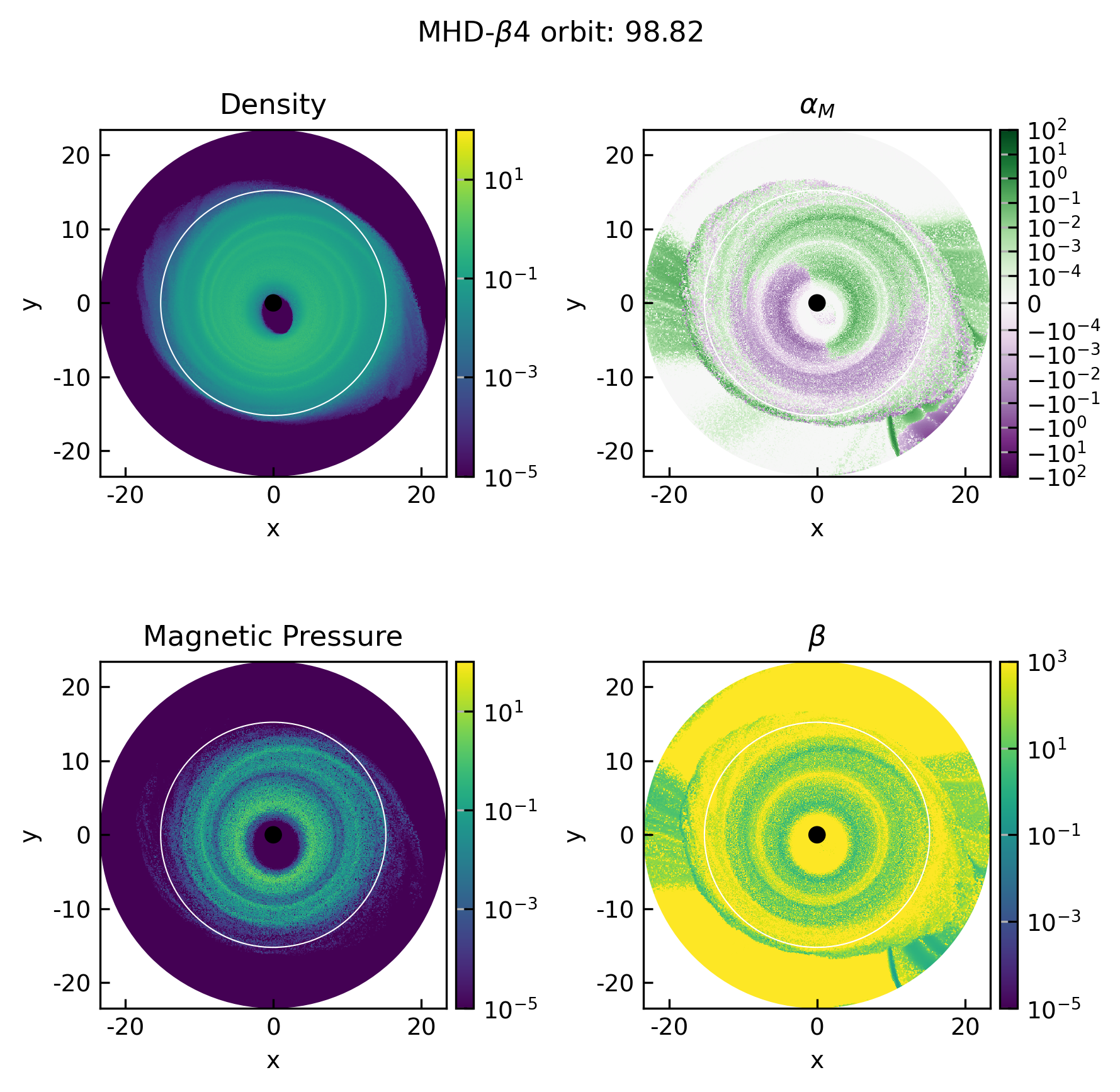}
    \centering
    \caption{Midplane ($z=0$) density (upper left), Maxwell stress alpha parameter (upper right), magnetic pressure (lower left), and plasma beta (lower right) in MHD-$\beta4$
    at the same epoch as shown in Figure~\ref{fig:late_time_1_and_11}.
    }
    \label{fig:late_time_alpha_beta_11}
\end{figure*}

In their simulations of eccentric disks, \citet{chan22} observed that the sign of the Maxwell stress flipped sign around the elliptical orbits, being positive for material falling inward toward periapsis, and negative for material moving outward toward apoapsis.  They provided a convincing explanation of this behavior from the changes in sign of radial velocity and its azimuthal derivative as one goes around an eccentric orbit, which affects the evolution of the sign of the Maxwell stress through flux-freezing.   We also observe Maxwell stresses to be positive on going from apoapsis to periapsis and negative on the return to apoapsis.  There nevertheless remains a net positive stress as must be the case for MRI turbulence to facilitate outward angular momentum transport. This splitting into positive and negative regions of stress is shown clearly in \autoref{fig:late_time_alpha_beta_11} where it is also once more apparent that the disk has broken and the inner and outer disks are misaligned.  This flipping of sign
of the Maxwell stress appears to be quite ordered between the inward and outward moving portions of the eccentric orbits in both the inner and outer eccentric disks.  This contrasts with the somewhat more disorderly behavior in the vertical field simulations of \citet{chan22} (compare with their Figure 5).  The more orderly behavior that we find is more similar to their dipole field simulations.  This difference between our vertical field simulations and theirs may be due to the fact that eccentricity grew from zero in our simulations, whereas their eccentricity was initialized with $e=0.5$, a value that our simulations never reach.

\begin{figure*}
    \includegraphics[width=18cm]{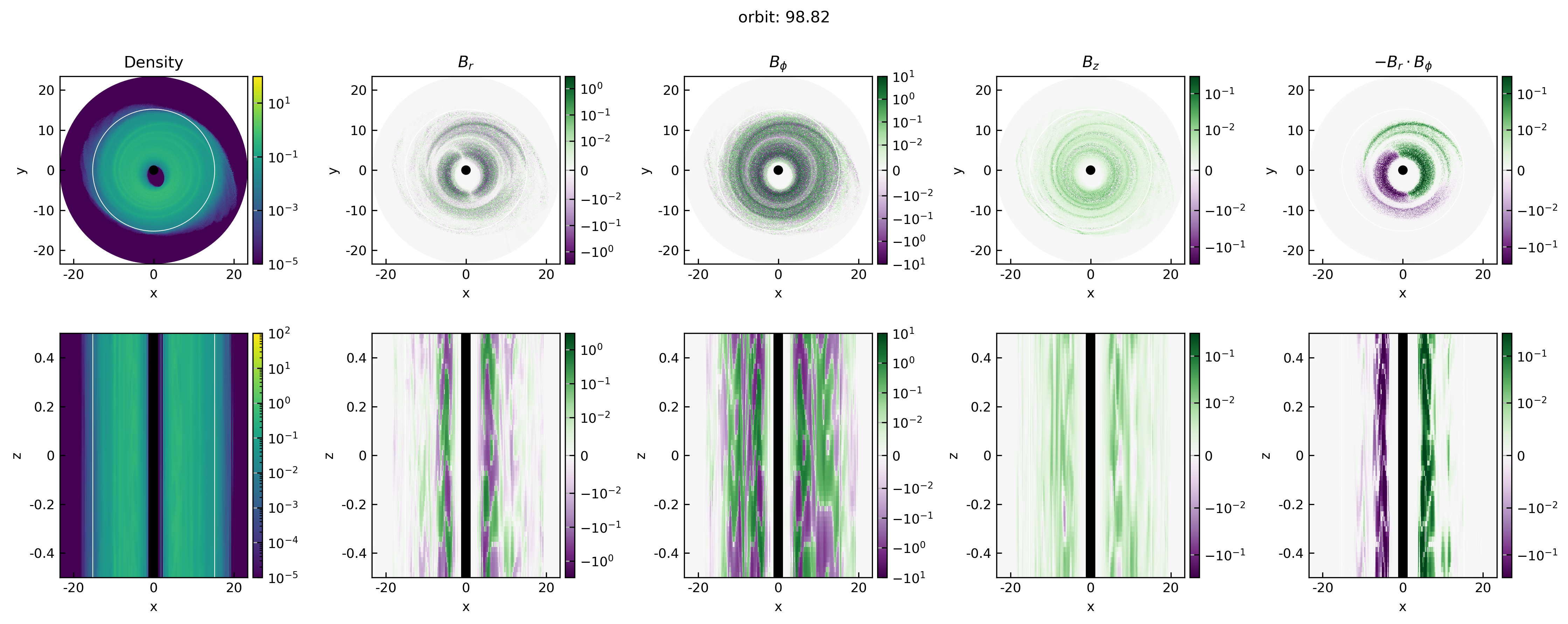}
    \caption{Late time snapshot of the magnetic field configuration of MHD-$\beta4$.
    The upper row shows $z=0$ midplane slices, and the bottom row shows $y=0$ vertical slices, of (left to right) density, the radial magnetic field component, the azimuthal magnetic field component, the vertical magnetic field component, and minus the product of the radial and azimuthal components (i.e. the Maxwell stress) to highlight their strong correlation despite large spatial variation in each independently.
    }
    \label{fig:late_time_field_11}
\end{figure*}

\begin{figure*}
    \includegraphics[width=16cm]{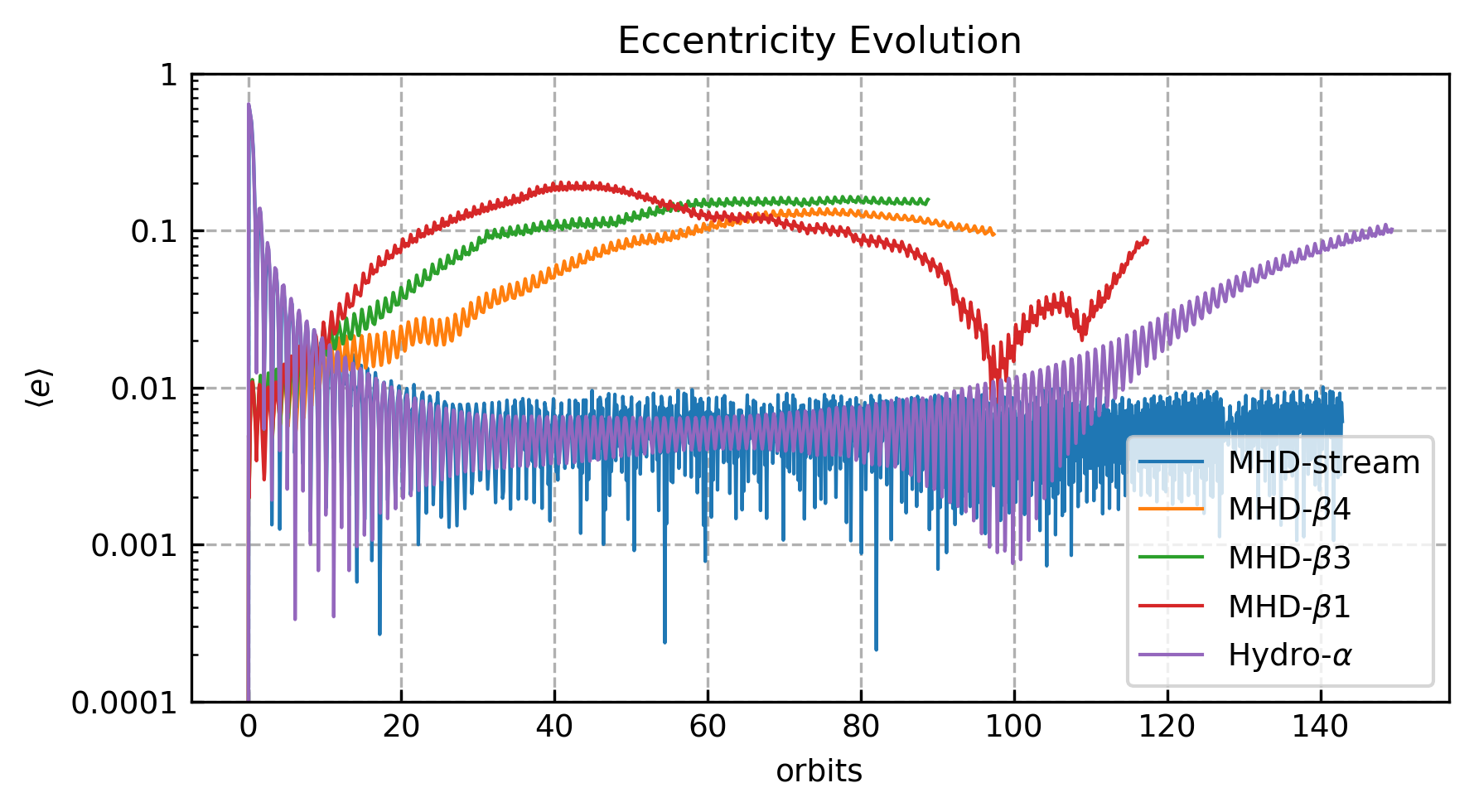}
    \centering
    \caption{Evolution of the magnitude of mass weighted average eccentricity across all the simulations.}
    \label{fig:eccent_comp}
\end{figure*}

\autoref{fig:late_time_alpha_beta_11} also reveals that the circular region that separates the misaligned inner and outer eccentric disks has much higher plasma $\beta$ due to much lower magnetic pressure there.  (Note that the density distribution is much
more smoothly varying through this region.)  This feature is not generic, and over the course of the simulation, we sometimes find strong variations in the plasma beta in the disk breaking region, due to variations in magnetic pressure or to density (and therefore gas pressure), or both.  The distributions of all three magnetic field components in the midplane and a vertical slice are shown in \autoref{fig:late_time_field_11}.  The radial and azimuthal magnetic field components have considerable small scale fluctuations in sign, but these fluctuations are correlated to give the same sign of stress on each side of the eccentric orbits, consistent with MRI turbulence.  The drop in magnetic pressure in the circular region separating the inner and outer eccentric disks arises from the drop in the magnitude of the radial and azimuthal field components in this region at this epoch, and this in turn results in a reduction of Maxwell stress.  Again, this behavior changes over the course of the simulation.

Turning now to the behavior of the other MHD simulations, \autoref{fig:eccent_comp} compares the global eccentricity evolution across all of them.  Eccentricity developed in simulations MHD-$\beta3$ and MHD-$\beta1$ 
in a very similar manner to MHD-$\beta4$: 
the disks broke into inner and outer misaligned regions with eccentric voids at the center.  However both the rate of eccentricity growth and the peak eccentricity increased monotonically with lower initial \(\beta\).  This is consistent with the fact that lower initial \(\beta\) results in higher \(\langle \alpha_M \rangle\), as shown in \autoref{fig:parameter_evolution}, and that the viscous hydrodynamical simulations of \citet{kle08} found that higher viscosity enhances the Lubow mechanism.  Hydro-$\alpha$ has an even larger alpha stress parameter than MHD-$\beta1$, MHD-$\beta3$, and MHD-$\beta4$, but its growth is delayed by the time it takes to spread the disk to the resonant radius, and the eccentricity growth may also be slowed by the presence of the stream \citep{lubow_eccentricity_damping_by_stream, kle08}.

MHD-$\beta1$,
which has the strongest initial magnetic field and strongest resulting MRI stresses, shows some interesting differences with MHD-$\beta4$ and MHD-$\beta3$.
At early times the tides quickly build up eccentricity in the outer regions, and the disk in fact breaks into three regions: a dense circular core, a highly eccentric outer disk, and between them a very diffuse region. This continues until about 40 orbits when the outer disk becomes so eccentric that its periapsis starts to make contact with the circular core. This triggers a contraction of the diffuse middle region as the outer disk spreads inward onto the core, forming a more homogeneous disk much like the early stages in the evolution of MHD-$\beta4$ and MHD-$\beta3$.
It is our opinion that this initial stage was a transient artifact of our initial condition where the magnetic pressure from the initialized magnetic annulus was strong enough to expand inward and outward, forming the low density diffuse region between the inner and outer disk.  As such it is the evolution after this structure has disappeared which is of most interest.

\begin{figure*}
    \includegraphics[width=9cm]{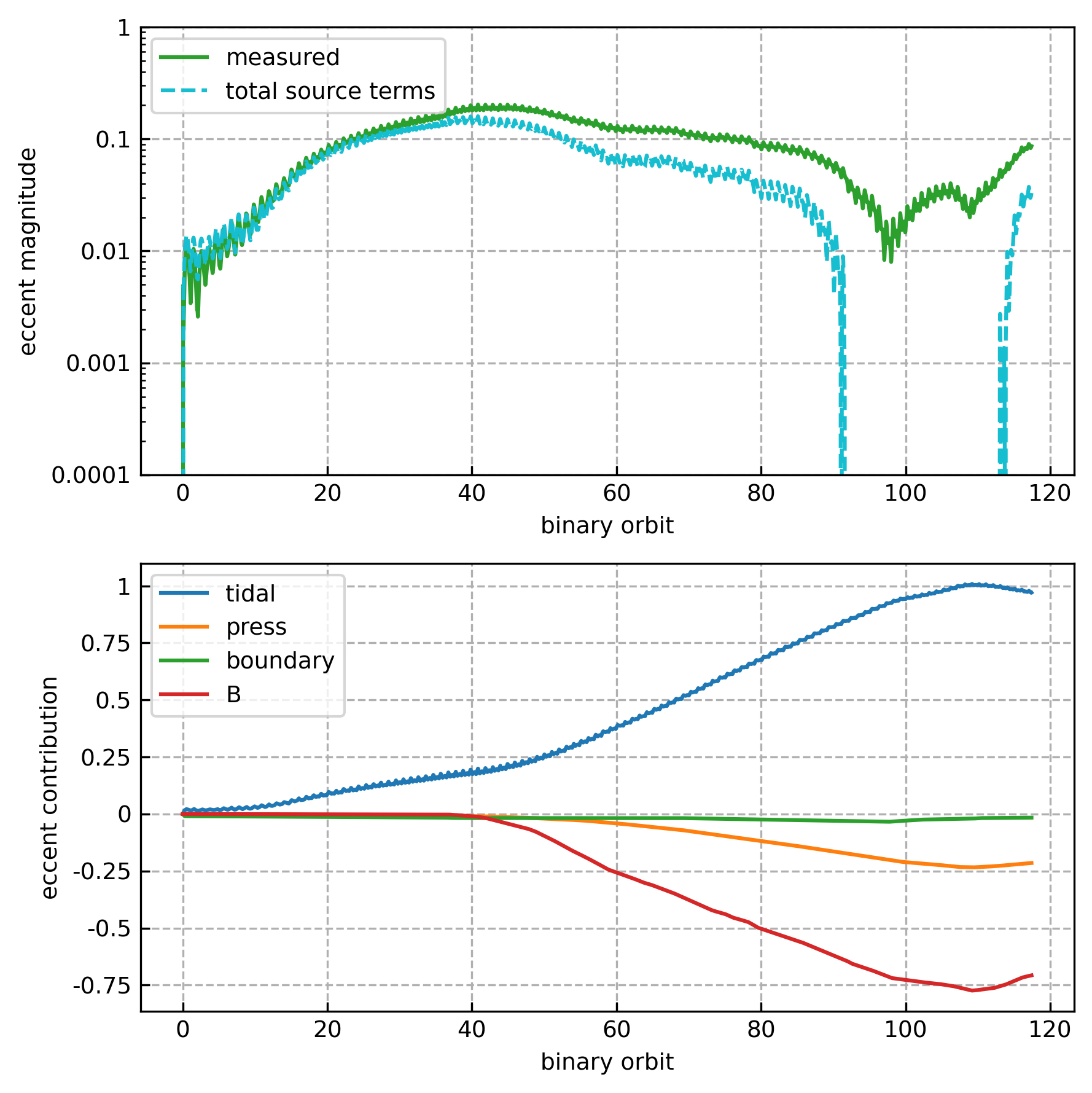}
    \includegraphics[width=9cm]{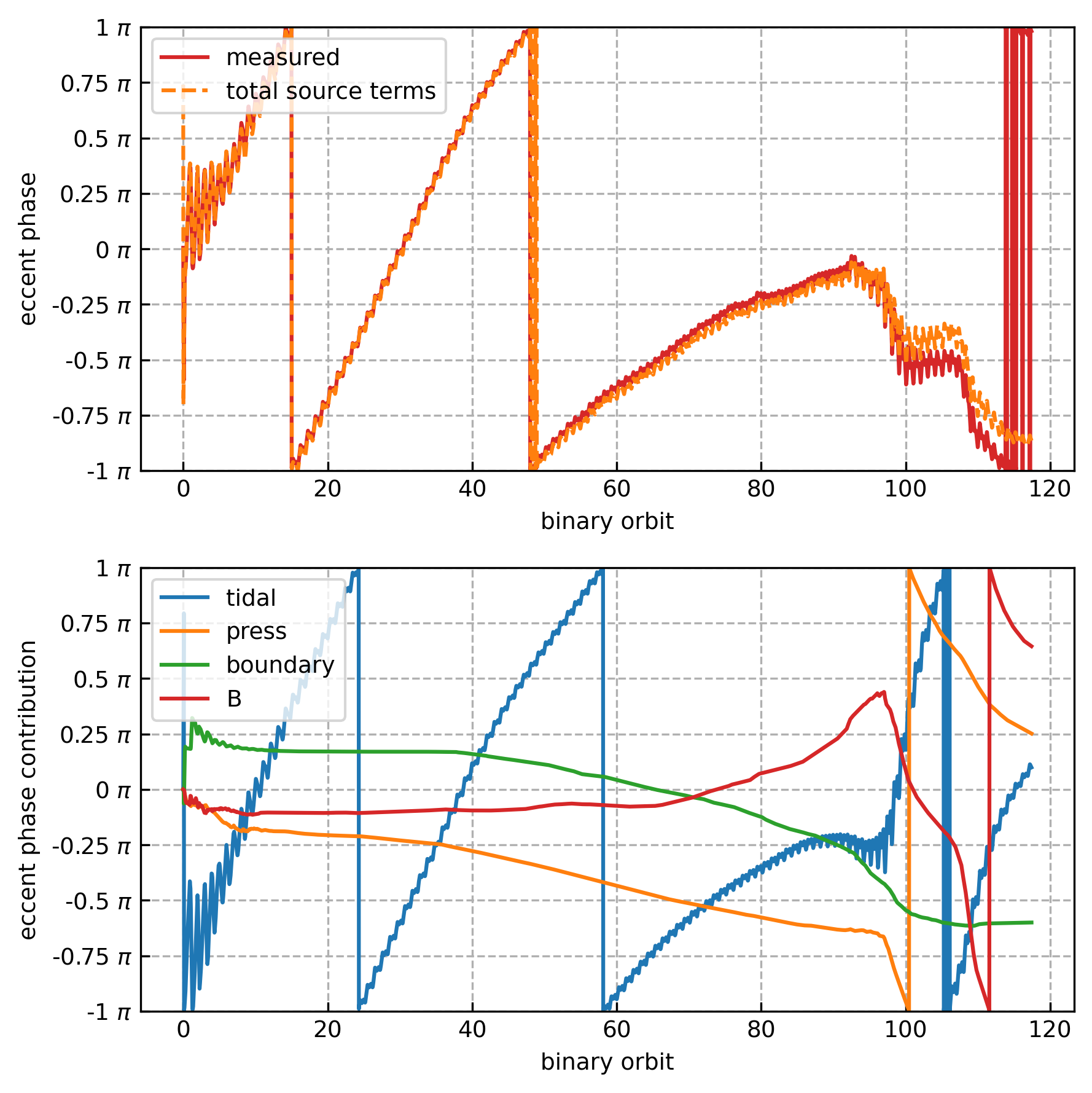}
    \caption{Evolution of eccentricity for simulation MHD-$\beta1$.
    As in \autoref{fig:Cyl_11_2_growth_prec} which was for the more weakly magnetized simulation MHD-$\beta4$,
    the left and right columns of panels refer to the eccentricity magnitude and direction in a non-rotating frame.  The top panels show the overall evolution, and the bottom panels show the cumulative contributions from various source terms.} 
    \label{fig:Cyl_15_2_growth_prec}
\end{figure*}

The inward spreading of the outer disk onto the inner core is also accompanied by substantial outward spreading of the outer disk, significantly increasing the strength of the tidal source terms in the eccentricity evolution. 
At the same time, \autoref{fig:Cyl_15_2_growth_prec} shows that MRI turbulence starts to act as a strong eccentricity sink, greater even than the enhanced tidal source.
The result of this is that the evolution from around orbit 40 to 100 looks very similar to MHD-$\beta4$ or MHD-$\beta3$
in that the disk breaks and an inner void forms in all the ways discussed previously.  However, unlike the weaker magnetic field simulations which began with low eccentricity that increased over time, MHD-$\beta1$
enters this regime with high eccentricity left over from the previous phase and slowly loses eccentricity throughout this epoch.  Notably it is only now when the inner and outer disk are in contact that the inner void starts to form and the disk breaks in the sense that the inner disk becomes highly eccentric and misaligned with the outer disk. The fact that all this happens during this epoch when the total eccentricity is declining strongly confirms the picture that these eccentric MRI disks are somehow relocating eccentricity from the outer disk to the inner disk. As the disk mass drains through the inner boundary, both the tidal and magnetic eccentricity source terms decrease in magnitude.  This continues until right around orbit 100 when they both pass through zero and flip sign, leading to a final epoch captured only briefly until the end of the simulation where tides act to damp eccentricity while the stronger magnetic source terms effectively cause it to grow (see \autoref{fig:Cyl_15_2_growth_prec}).\footnote{Note that the sum of the source terms in \autoref{fig:Cyl_15_2_growth_prec} has a small accumulating error that causes it to go very slightly negative during these times, which is why it drops off this logarithmic plot.  The accumulating error between the measured eccentricity and the sum of the source terms is larger here than in simulation MHD-$\beta4$ (cf. \autoref{fig:Cyl_11_2_growth_prec}), likely due to the faster evolution in MHD-$\beta1$ and our finite time sampling.  A similar plot of the eccentricity evolution of MHD-$\beta3$ shows a small accumulating error between that of MHD-$\beta1$ and MHD-$\beta4$, presumably due to its intermediate rate of evolution (cf. \autoref{fig:eccent_comp}).}

We also see that at this turnover moment at $\simeq90$ orbits, the magnetic source terms, which had previously been driving mild prograde precession, suddenly flip to be an immense source of retrograde precession. Meanwhile tides (which generally drive prograde precession) and gas pressure (which is driving retrograde precession), also dramatically increase in magnitude.  This sudden transition results in the disk actually halting its prograde precession and moving into a retrograde phase.  Unlike the transition in behaviour around orbit 40, the cause of this transition around orbit 90 is unclear. Despite how dramatic the transition appears in the source terms nothing particularly dramatic is seen in visualizations of the density and eccentricity.

Hydrodynamical eccentric disks experience a parametric instability that excites inertial waves and damps eccentricity \citep{pap05a,bar14}.  Simulations of this in vertically unstratified accretion disks produce patterns of vertical oscillation that vary in both radius and height \citep{pap05b}.  We find no evidence of the presence of such oscillations in any of our simulations that developed eccentricity.  Indeed, very little vertical kinetic energy is present, despite the presence of vertical gradients of magnetic field components as shown, for example, in \autoref{fig:late_time_field_11}.  This lack of vertical motion is present even in Hydro-$\alpha$, which shows that MRI turbulence itself is not the cause of suppression of inertial waves in our simulations.  It is unclear as to why the parametric instability is not present, unless alpha-viscosity itself can suppress it.  It may simply be that we have not run our simulations for long enough.  For small eccentricities, the instability growth rate is $3e\omega/16$, where $\omega$ is the local disk orbital frequency \citep{pap05a}.  Hence the growth rate times the binary orbital period ranges from $\simeq220e$ near the inner boundary to $\simeq4e$ near the resonant radius.  One hundred binary orbital periods at $e=0.1$ therefore only gives us at most five e-foldings of growth.

\section{Discussion}
\label{sec:discussion}

Perhaps the most central result of this paper is that while it is true that MRI turbulence acts as an eccentricity sink, once at the resonant radius this effect is easily overpowered by the tidal forces which source eccentricity. One might be inclined to think MRI turbulence is an eccentricity sink everywhere because it is a dissipative process and as such should circularize orbits, but remarkably this is not the case. While it is globally an eccentricity sink because it is dissipative and thus acts to globally circularize the disk, it can act locally as an eccentricity source and indeed we see that it does. In both our simulation and those of \citet{chan22} we see the MRI favors moving eccentricity inwards. 
This leads to the amplification of eccentricity in the inner regions of the disk with a tendency toward circularization of the outer regions. \citet{chan22} have shown that the primary driver of this is that MRI turbulence preferentially transports angular momentum over energy, which would naturally circularize outer orbits and pump eccentricity inward. In our case, since the Lubow mechanism provides a constant source of eccentricity at the resonant radius, we developed a highly eccentric inner disk with eccentric standing waves \citep{ogilvie_lynch_2018}, surrounded by an outer disk where eccentric waves continue to be launched at the resonant radius. It is also worth noting that this process in which eccentricity grows in the inner disk at the expense of eccentricity in the outer disk continues well after the eccentricity of the disk has saturated. In fact in MHD-\(\beta1\) this process begins only after the disk has already reached its maximum eccentricity and begun to decline. During this process, the inner and outer regions become misaligned in eccentricity, and the region between them consists of circular orbits.
The most likely candidate for a circularization process in this intermediate region are collisions between the misaligned eccentric orbits in the inner and outer regions.  
In other words, we have two distinct eccentric precessing inner and outer disks which are shearing against each other at a circular boundary. This boundary stands out prominently in the top left panel of \autoref{fig:late_time_1_and_11}.

The differential precession between these inner and outer eccentric disks is likely to result in a broader spectrum of observed frequencies due to the presence of two apsidal precession periods rather than just one.  It is therefore perhaps noteworthy that our simulation parameters were modeled after the AM~CVn system SDSS J1908 \citep{fon11} which has a rich spectrum of observed frequencies, some of which are more stable in time than others \citep{kup15}.

Another interesting feature of our simulations is the formation of eccentric inner voids. The structure of the inner disk seems to be a series of eccentric standing waves such as those described in \citet{ogilvie_lynch_2018}, bounded inward by an eccentric ring with periapsis right on the edge of the inner boundary of the simulation domain. Inside of this ring we have a near total void. Any portions of the standing wave that would have existed in this void region would have their periapses inside the inner boundary. With our inflow boundary conditions this of course means that material inside the edge of the void plunge inward and are removed from the simulation.  It is likely that replacing the inflow boundary condition with a stellar boundary layer would affect the formation of the void, and possibly the entire standing wave pattern in the inner disk.  Indeed, the viscous hydrodynamical simulations of \citet{kle08} also produced eccentric inner voids only when they employed an inflow inner radial boundary condition.  More work therefore needs to be done to understand the effects of realistic inner boundary conditions on the structure of the inner disk.  If standing waves persist, it would be interesting to investigate what luminosity variability they might produce.

Generally we see the Lubow mechanism excite eccentricity at the resonance radius as in alpha disk simulations.  However, eccentricity in MHD disks congregates in the innermost orbits leading to large eccentric voids, a result also observed in \citet{chan22}. Specifically, they observed that a vertical field configuration did not lead to an obvious eccentricity flow but a dipole field drove a clear flow of eccentricity from the outer disk in. This is somewhat in contrast to our results which all used vertical fields, however we ran our simulations for about a hundred binary orbits while \citep{chan22} only ran theirs for 15 orbital periods of their initial eccentric disk annulus.   It is therefore possible that this effect is much weaker in vertical field configurations and thus only made itself noticeable in longer simulation runs. Additionally we observed the nonaxisymmetric Maxwell stress around the orbit as noted by \citet{chan22}, where the negative stresses occur on the outflowing side and positive stresses on the inflowing side.

Our primary result that the Lubow mechanism remains robust in the presence of MRI turbulence is not without caveats. We neglected vertical stratification which would introduce further complexity such as the vertical breathing modes described in \citet{chan23}. Additionally we included no interactions with the white dwarf at the boundary which in reality may have affected the evolution of the inner voids we observed. In either extreme whether material inelastically sticks to the white dwarf where it impacts immediately dissipating all its energy creating a hot spot or if material brushes the surface many times slowly circularizing until it comes to rest on the surface, we can be sure the innermost region has notable features we have neglected in this work. We also neglected realistic thermodynamics, instead just enforcing a locally isothermal equation of state for our plasma.  Since we believe the circularization of the region between the inner and outer disk is due to dissipative orbit crossings, we would expect to see enhanced heating in this region.  It would be interesting to explore this with vertically stratified simulations with more realistic thermodynamics.

Lastly and perhaps most importantly, we did not include an accretion stream in the MHD simulations that went eccentric. A key result of this paper is our realization that building up a disk from a stream and have it spread out to the resonant radius is a nontrivial exercise in the presence of MRI turbulence.  How exactly this is accomplished in Nature is an important puzzle we have only highlighted but not solved.  It may be that this is coupled with our neglect of realistic thermodynamics, as the formation of dense rings might be affected by our locally isothermal treatment.  Build-up of matter in an outer, cool dense ring that is fed by a stream may eventually trigger an ionization instability, triggering MRI and an outburst.  This might partially explain why positive superhumps are so strongly associated with superoutbursts in SU~UMa type systems.  However, this does not answer the question as to how persistent stream-fed systems manage to maintain superhumps.
Note that \citet{kle08} found that the presence of a stream enabled them to obtain prograde apsidal precession rates in their viscous hydrodynamical simulations that were in line with observations.  This again emphasizes the need to understand stream-fed systems in the presence of MRI turbulence.

An important caveat concerning the failure of MHD-stream to spread to the resonant radius is that the MRI can only start to grow once the stream completes an orbit and forms a ring at the circularization radius.  After that, the MRI will enable fluctuations to e-fold on approximately the local orbital time, but it will still generally take many of those times to reach a state of saturated turbulence.  In the meantime, mass is still being added by the stream, growing the density of the ring on a dynamical time.  Hence once the disk starts to spread, it may do so with an overdense outer edge which gives tidal torques an advantage over at least the Maxwell stresses within MRI turbulence.  It might therefore be an interesting future numerical experiment to begin with a ring at the circularization radius that already has fully-developed turbulence that is then fed by a stream.

\section{Conclusions}
\label{sec:conclusions}

We have shown that MHD disks which have spread to the resonant radius will develop eccentricities via the Lubow mechanism and will do so on time scales comparable to those seen in simulations of alpha disks. MRI turbulence is therefore entirely compatible with the Lubow mechanism for resonant growth of accretion disk eccentricity in compact binary systems.  However, MRI turbulence interacts with eccentricity in interesting ways that differ from alpha disks. MRI turbulence 
favors moving eccentricity towards the center of the disk leading to highly eccentric standing waves and voids which break, in terms of eccentricity orientation, from the outer disk.   We do not see such behavior in the alpha-disk simulation, even though it becomes eccentric.  For the MHD eccentric disks, the outer region hosts waves that are excited at the resonant radius and carry eccentricity inward to a boundary region between the inner and outer eccentric disks. The boundary region, despite being surrounded by an eccentric disk, remains circular.

Our secondary result is that spreading disks in the presence of MRI is nontrivial. In an alpha disk model the ratio between viscous and tidal torques is approximately independent of disk structure, so disk truncation is largely determined by the tidal potential and is less affected by how the disk evolves. In the presence of MRI turbulence, this is not the case. We demonstrated in the MHD-stream simulation how naive attempts to build a disk up from a spreading ring at the circularization radius can result in early truncation due to the formation of over-dense rings where tides overpower MRI stresses.  This was true even though MHD-stream had the highest alpha stress parameters ($\langle\alpha_M\rangle=0.077$, $\langle\alpha_R\rangle=0.091$) of any of our MHD simulations, and higher total stress parameter than our stream-fed Hydro-$\alpha$ simulation $(\alpha=0.1$), which did successfully spread to the resonant radius and grow eccentricity.   If, on the other hand, a disk is established out at the resonant radius, even relatively weak turbulent stresses ($\langle\alpha\rangle\sim0.024$ in the case of MHD-$\beta$4) can maintain it there.

\begin{acknowledgments}

We thank Janosz Dewberry, Jonatan Jacquemin-Ide, Steve Lubow, and John Papaloizou for useful conversations.  This work was supported in part by NASA Astrophysics Theory Program grant 80NSSC20K0525, and by grant NSF PHY-2309135 to the Kavli Institute for Theoretical Physics (KITP).  Computational resources were provided by the NASA High-End Computing (HEC) Program through the
NASA Advanced Supercomputing (NAS) Division at Ames Research Center.  Additional computational facilities used in this research were purchased with funds from the National Science Foundation (CNS-1725797) and administered by the Center for Scientific Computing (CSC). The CSC is supported by the California NanoSystems Institute and the Materials Research Science and Engineering Center (MRSEC; NSF DMR 1720256) at UC Santa Barbara. The Center for Computational Astrophysics at the Flatiron Institute is supported by the Simons Foundation.
\end{acknowledgments}

\bibliography{references}{}

\begin{thebibliography}{}
\expandafter\ifx\csname natexlab\endcsname\relax\def\natexlab#1{#1}\fi
\providecommand{\url}[1]{\href{#1}{#1}}
\providecommand{\dodoi}[1]{doi:~\href{http://doi.org/#1}{\nolinkurl{#1}}}
\providecommand{\doeprint}[1]{\href{http://ascl.net/#1}{\nolinkurl{http://ascl.net/#1}}}
\providecommand{\doarXiv}[1]{\href{https://arxiv.org/abs/#1}{\nolinkurl{https://arxiv.org/abs/#1}}}

\bibitem[{{Balbus} \& {Hawley}(1998)}]{bal98}
{Balbus}, S.~A., \& {Hawley}, J.~F. 1998, Reviews of Modern Physics, 70, 1,
  \dodoi{10.1103/RevModPhys.70.1}

\bibitem[{{Barker} \& {Ogilvie}(2014)}]{bar14}
{Barker}, A.~J., \& {Ogilvie}, G.~I. 2014, \mnras, 445, 2637,
  \dodoi{10.1093/mnras/stu1939}

\bibitem[{{Chan} {et~al.}(2018){Chan}, {Krolik}, \& {Piran}}]{chan18}
{Chan}, C.-H., {Krolik}, J.~H., \& {Piran}, T. 2018, \apj, 856, 12,
  \dodoi{10.3847/1538-4357/aab15c}

\bibitem[{{Chan} {et~al.}(2022){Chan}, {Piran}, \& {Krolik}}]{chan22}
{Chan}, C.-H., {Piran}, T., \& {Krolik}, J.~H. 2022, \apj, 933, 81,
  \dodoi{10.3847/1538-4357/ac68f3}

\bibitem[{{Chan} {et~al.}(2023){Chan}, {Piran}, \& {Krolik}}]{chan23}
---. 2023, arXiv e-prints, arXiv:2312.06775, \dodoi{10.48550/arXiv.2312.06775}

\bibitem[{{Dewberry} {et~al.}(2020){Dewberry}, {Latter}, {Ogilvie}, \&
  {Fromang}}]{dew20}
{Dewberry}, J.~W., {Latter}, H.~N., {Ogilvie}, G.~I., \& {Fromang}, S. 2020,
  \mnras, 497, 451, \dodoi{10.1093/mnras/staa1898}

\bibitem[{{Fontaine} {et~al.}(2011){Fontaine}, {Brassard}, {Green},
  {Charpinet}, {Dufour}, {Hubeny}, {Steeghs}, {Aerts}, {Randall}, {Bergeron},
  {Guvenen}, {O'Malley}, {Van Grootel}, {{\O}stensen}, {Bloemen}, {Silvotti},
  {Howell}, {Baran}, {Kepler}, {Marsh}, {Montgomery}, {Oreiro}, {Provencal},
  {Telting}, {Winget}, {Zima}, {Christensen-Dalsgaard}, \& {Kjeldsen}}]{fon11}
{Fontaine}, G., {Brassard}, P., {Green}, E.~M., {et~al.} 2011, \apj, 726, 92,
  \dodoi{10.1088/0004-637X/726/2/92}

\bibitem[{{Fromang} {et~al.}(2013){Fromang}, {Latter}, {Lesur}, \&
  {Ogilvie}}]{fro13}
{Fromang}, S., {Latter}, H., {Lesur}, G., \& {Ogilvie}, G.~I. 2013, \aap, 552,
  A71, \dodoi{10.1051/0004-6361/201220016}

\bibitem[{{Kley} {et~al.}(2008){Kley}, {Papaloizou}, \& {Ogilvie}}]{kle08}
{Kley}, W., {Papaloizou}, J.~C.~B., \& {Ogilvie}, G.~I. 2008, \aap, 487, 671,
  \dodoi{10.1051/0004-6361:200809953}

\bibitem[{{Kunze} {et~al.}(1997){Kunze}, {Speith}, \& {Riffert}}]{kun97}
{Kunze}, S., {Speith}, R., \& {Riffert}, H. 1997, \mnras, 289, 889,
  \dodoi{10.1093/mnras/289.4.889}

\bibitem[{{Kupfer} {et~al.}(2015){Kupfer}, {Groot}, {Bloemen}, {Levitan},
  {Steeghs}, {Marsh}, {Rutten}, {Nelemans}, {Prince}, {F{\"u}rst}, \&
  {Geier}}]{kup15}
{Kupfer}, T., {Groot}, P.~J., {Bloemen}, S., {et~al.} 2015, \mnras, 453, 483,
  \dodoi{10.1093/mnras/stv1609}

\bibitem[{{Lubow}(1991)}]{lubow_theory}
{Lubow}, S.~H. 1991, \apj, 381, 259, \dodoi{10.1086/170647}

\bibitem[{{Lubow}(1994)}]{lubow_eccentricity_damping_by_stream}
---. 1994, \apj, 432, 224, \dodoi{10.1086/174563}

\bibitem[{{Lynch} \& {Dewberry}(2023)}]{lyn23}
{Lynch}, E.~M., \& {Dewberry}, J.~W. 2023, \mnras, 526, 2673,
  \dodoi{10.1093/mnras/stad2678}

\bibitem[{{Murray}(1998)}]{mur98}
{Murray}, J.~R. 1998, \mnras, 297, 323,
  \dodoi{10.1046/j.1365-8711.1998.01504.x}

\bibitem[{{O'Donoghue} \& {Charles}(1996)}]{odo96}
{O'Donoghue}, D., \& {Charles}, P.~A. 1996, \mnras, 282, 191,
  \dodoi{10.1093/mnras/282.1.191}

\bibitem[{Ogilvie \& Lynch(2018)}]{ogilvie_lynch_2018}
Ogilvie, G.~I., \& Lynch, E.~M. 2018, Monthly Notices of the Royal Astronomical
  Society, 483, 4453, \dodoi{10.1093/mnras/sty3436}

\bibitem[{{Oyang} {et~al.}(2021){Oyang}, {Jiang}, \& {Blaes}}]{Oyang2021}
{Oyang}, B., {Jiang}, Y.-F., \& {Blaes}, O. 2021, \mnras, 505, 1,
  \dodoi{10.1093/mnras/stab1212}

\bibitem[{{Papaloizou}(2005{\natexlab{a}})}]{pap05a}
{Papaloizou}, J.~C.~B. 2005{\natexlab{a}}, \aap, 432, 743,
  \dodoi{10.1051/0004-6361:20041947}

\bibitem[{{Papaloizou}(2005{\natexlab{b}})}]{pap05b}
---. 2005{\natexlab{b}}, \aap, 432, 757, \dodoi{10.1051/0004-6361:20041948}

\bibitem[{{Patterson} {et~al.}(1993){Patterson}, {Thomas}, {Skillman}, \&
  {Diaz}}]{pat93}
{Patterson}, J., {Thomas}, G., {Skillman}, D.~R., \& {Diaz}, M. 1993, \apjs,
  86, 235, \dodoi{10.1086/191777}

\bibitem[{{Shakura} \& {Sunyaev}(1973)}]{sha73}
{Shakura}, N.~I., \& {Sunyaev}, R.~A. 1973, \aap, 24, 337

\bibitem[{{Stone} {et~al.}(2020){Stone}, {Tomida}, {White}, \&
  {Felker}}]{athena}
{Stone}, J.~M., {Tomida}, K., {White}, C.~J., \& {Felker}, K.~G. 2020, \apjs,
  249, 4, \dodoi{10.3847/1538-4365/ab929b}

\bibitem[{{Suzuki} \& {Inutsuka}(2009)}]{suz09}
{Suzuki}, T.~K., \& {Inutsuka}, S.-i. 2009, \apjl, 691, L49,
  \dodoi{10.1088/0004-637X/691/1/L49}

\bibitem[{{Vogt}(1974)}]{vogt74}
{Vogt}, N. 1974, \aap, 36, 369

\bibitem[{{Warner}(1975)}]{war75}
{Warner}, B. 1975, \mnras, 170, 219, \dodoi{10.1093/mnras/170.1.219}

\bibitem[{{Whitehurst}(1988)}]{whi88}
{Whitehurst}, R. 1988, \mnras, 232, 35, \dodoi{10.1093/mnras/232.1.35}

\bibitem[{{Whitehurst}(1994)}]{whi94}
---. 1994, \mnras, 266, 35, \dodoi{10.1093/mnras/266.1.35}

\bibitem[{{Whitehurst} \& {King}(1991)}]{whi91}
{Whitehurst}, R., \& {King}, A. 1991, \mnras, 249, 25,
  \dodoi{10.1093/mnras/249.1.25}

\bibitem[{{Zurita} {et~al.}(2002){Zurita}, {Casares}, {Shahbaz}, {Wagner},
  {Foltz}, {Rodr{\'\i}guez-Gil}, {Hynes}, {Charles}, {Ryan}, {Schwarz}, \&
  {Starrfield}}]{zur02}
{Zurita}, C., {Casares}, J., {Shahbaz}, T., {et~al.} 2002, \mnras, 333, 791,
  \dodoi{10.1046/j.1365-8711.2002.05450.x}

\end{thebibliography}
\bibliographystyle{aasjournal}



\end{CJK*}
\end{document}